\documentclass[manuscript]{aastex631}

\usepackage{newtxtext,newtxmath}
\usepackage[T1]{fontenc}
\usepackage{graphicx}
\usepackage{amsmath}
\usepackage{lipsum}
\usepackage{hyperref}
\begin{document}

\title{Properties of Voids and Void Galaxies in the TNG300 Simulation}

\author[0000-0002-0212-4563]{Olivia Curtis}
\affiliation{Department of Astronomy \& Institute for Astrophysical Research, 725 Commonwealth Ave., Boston University, Boston, MA 02215, USA}

\author[0000-0001-6928-4345]{Bryanne McDonough}
\affiliation{Department of Astronomy \& Institute for Astrophysical Research, 725 Commonwealth Ave., Boston University, Boston, MA 02215, USA}

\author[0000-0001-7917-7623]{Tereasa G. Brainerd}
\affiliation{Department of Astronomy \& Institute for Astrophysical Research, 725 Commonwealth Ave., Boston University, Boston, MA 02215, USA}

% Abstract of the paper
\begin{abstract}
%
% The abstract should not contain much in the way of motivation.  Instead, it needs to
% focus on what you did and what you found
%
% Before you talk about the radial density profile as defined by void galaxies, you need
% to talk about identifying void galaxies.
%
% ApJ limits the length of the abstract to 250 words
%
We investigate the properties of voids and void galaxies in the \texttt{TNG300} simulation. Using a luminous galaxy catalog and a spherical void finding algorithm,
we identify 5,078 voids at redshift $z = 0$.
%\textcolor{red}{The Magneticum midres simulation has > 300,000 voids defined using halos; hires has > 33,000 voids defined using halos}
The voids cover 83\% of the simulation volume and have a median radius of $4.4h^{-1} \rm{Mpc}$. 
We identify two populations of field galaxies based on whether the galaxies reside within a void (``void galaxies''; 75,220 objects) or outside a void (``non-void galaxies''; 527,454 objects). 
Within the voids, mass does not directly trace light.  Instead, the mean radial underdensity profile as defined by the locations of void galaxies is systematically lower than the mean radial underdensity profile as defined by the dark matter (i.e., the voids are more ``devoid'' of galaxies than they are of mass).  
Within the voids, the integrated underdensity profiles of the dark matter and the galaxies are independent of the local background density (i.e., voids-in-voids vs.\ voids-in-clouds).  Beyond the void radii, however, the integrated underdensity profiles of both the dark matter and the galaxies exhibit strong dependencies on the local background density.
Compared to non-void galaxies, void galaxies are on average younger, less massive, bluer in color, less metal enriched, and have smaller radii. 
In addition, the specific star formation rates of void galaxies are $\sim 20$\% higher than non-void galaxies and, in the case of galaxies with central supermassive black holes with $M_{\rm BH} \gtrsim 3\times 10^6 h^{-1} M_\odot$, the fraction of active void galaxies is $\sim 25$\% higher than active non-void galaxies.
\end{abstract}

\keywords{Large-scale structure of the Universe (902) -- Voids (1779) -- Magnetohydrodynamical simulations (1966) -- Galaxy evolution (594)}

%%%%%%%%%%%%%%%%% BODY OF PAPER %%%%%%%%%%%%%%%%%%

\section{Introduction}\label{sec:intro}

%\textcolor{red}{Need to incorporate discussion of Schuster et al. https://arxiv.org/pdf/2210.02457.pdf somewhere - perhaps in the Summary and Discussion section?} \textcolor{orange}{OC notes: Done, see line 437. FYI, the orange texts are general notes to you that will be removed and the blue texts are to be incorporated into the manuscript.}

The large-scale structure of the Universe is an interconnected network of walls, sheets, filaments, and galaxy clusters, between which lie vast regions of near nothingness known as cosmic voids (see, e.g., \citealt{giovanelli1991}). The origin of the ``cosmic web'' dates back to the early {Universe} during the epoch of inflation, when quantum fluctuations near the end of the inflationary period gave rise to slight anisotropies in the post-inflation matter density field. These anisotropies are observed in the cosmic microwave background (CMB), with regions that deviate by a few micro-Kelvin from the average CMB temperature (see, e.g., \citealt{planck2020}), and they are responsible for the structure that is seen in the cosmic web today.

Perturbations in the density field are unstable. Regions that are overdense experience an inward gravitational force, causing them to increase in density as they collapse and accrete surrounding matter \citep{bertschinger1998}. The opposite is true for regions that are initially underdense. Matter in these regions experiences an outward gravitational force that attracts it towards nearby, higher density regions. As matter streams out, these regions become more underdense, increasing the relative outward gravitational pull and causing them to expand even further. Since these expanding regions never become virialized, their growth can always be modelled by linear perturbation theory (see, e.g., \citealt{goldberg2004}).
Primordial overdensities in the post-inflation matter density field evolved in a bottom-up, hierarchical fashion consistent with Cold Dark Matter (CDM) theory (see, e.g., \citealt{liddle1993}), forming halos, clusters, walls, and filaments. Conversely, primordial underdensities expanded, becoming the voids.

Voids have properties that make them interesting regions for performing various cosmological tests \citep{sheth}, and the interiors of mature voids can be described as low density Friedmann-Lemaître-Robertson-Walker universes (see, e.g., \citealt{Icke}; \citealt{vandeweygaert1993}).  %Therefore, the interiors of voids can be thought of as miniature \textcolor{red}{Universe}s. 
This makes voids excellent laboratories for tests of the expansion rate and geometry of the {Universe}, the dark energy equation of state, and modified theories of gravity (see, e.g., \citealt{li2012}; \citealt{clampitt2013}; \citealt{gibbons2014}; \citealt{cai2015}; \citealt{pollina2016}; \citealt{cai2017}; \citealt{falck2018}; \citealt{paillas2019}). Furthermore, since voids are smaller than the mean-free path of neutrinos \citep{lesgourgues2006}, voids can be used to constrain the sum of neutrino masses via tests that examine the ways in which neutrino properties affect the sizes and distributions of voids (see, e.g., \citealt{villaescusanavarro2013}; \citealt{massara2015}; \citealt{banerjee2016}; \citealt{kreisch2019}; \citealt{schuster2019}; \citealt{contarini2021}).

Voids are not completely devoid of internal structure, and even the most mature voids in the local {Universe} show substructure in the form of galaxies and diffuse filaments \citep{szomoru1996, elad1997, hoyle2004, sheth, kreckel2012, alpaslan2014}. The dynamical structure within voids can have significant impact on cosmic flow patterns in the local {Universe}, with matter around the ridges of voids affecting the peculiar motions of galaxies around nearby walls and filaments (see, e.g., \citealt{bothun1992}; \citealt{vandeweygaert2016}; \citealt{vallesperez2021}; \citealt{bermejo2022}).  

Additionally, the fact that voids resemble low-density universes with large Hubble parameters (see, e.g., \citealt{Icke}; \citealt{goldberg2004}) makes void galaxies interesting objects with which to test models of galaxy formation and evolution. From linear theory, it is known that dark matter halos in underdense regions of the {Universe} form later in the history of the {Universe} than do their counterparts in overdense regions \citep{liddle1993}. Therefore, studying the physical properties of void galaxies in the local {Universe} has the potential to provide insight into an early epoch of galaxy evolution. 

Furthermore, the degree to which the location of a galaxy within the cosmic web affects the evolution of the galaxy is an open question.  Some studies have concluded that, due to their differing dynamical histories,  
void galaxies differ systematically from non-void field galaxies (see, e.g., \citealt{peebles2001}; 
\citealt{croton2005}; \citealt{kreckel2012}; \citealt{rodriguez2022}; \citealt{rosasguevara2022}). For example, when compared to observed galaxies in walls and and filaments, some studies have found that void galaxies are, on average, bluer in color, have lower stellar masses, are richer in HI, have higher specific star formation rates, and have later morphological types (see, e.g., \citealt{rojas2004};
\citealt{croton2005}; \citealt{hoyle2012}; \citealt{kreckel2012}; \citealt{beygu2016}; 
\citealt{douglass2018}; 
\citealt{florez2021}; 
\citealt{pandey2021};  \citealt{rodriguez2022}). However, other studies have found little to no difference between the stellar masses, gas content, star formation rates, chemical abundances, dark matter profiles, or metallicities of void vs.\ non-void field galaxies (see, e.g., \citealt{szomoru1996}; \citealt{patiri2006}; 
\citealt{moorman2014};
\citealt{liu2015};
\citealt{douglass2017};
\citealt{douglass2019}; \citealt{wegner2019}; \citealt{dominguezgomez2022}).

Properties of void and non-void galaxies have also been studied in recent magnetohydroynamical simulations of $\Lambda$CDM universes.  In agreement with some observational studies, \cite{rosasguevara2022} found that, compared to galaxies in denser environments, void galaxies in the \texttt{EAGLE} simulation \citep{schaye2015} have lower stellar mass fractions.  In addition, \cite{rosasguevara2022} found clear trends of galaxy properties as a function of the distances of galaxies from the nearest void center. In particular, star formation activity and HI gas density decreased with increasing void-centric distance, and stellar mass fraction increased with increasing void-centric distance.
%both star formation activity and HI abundance within a galaxy decrease with its distance to the closest void. 
%\textcolor{red}{Is the previous sentence a correct statement?} \textcolor{orange}{OC notes: Yes, this is correct.} 
Similarly, in a study of galaxies in the \texttt{HorizonAGN} simulation \citep{dubois2016}, \cite{habouzit} found that low stellar mass galaxies with high star formation rates occur more frequently in the inner regions of voids than in denser regions of the cosmic web.

Another open question involves the degree to which the location of a galaxy within the cosmic web influences the formation of active galactic nuclei (AGN). Some studies have concluded that the AGN fraction of galaxies is independent of the local matter density (e.g.,  \citealt{karhunen2014}; \citealt{sabater2015}; \citealt{amiri2019}; \citealt{habouzit}), while others have found a positive correlation with local matter density (e.g., \citealt{manzer2014}; \citealt{argudofern2018}). Still other studies have concluded that there is a negative correlation between AGN fraction and the local matter density (e.g., \citealt{kauffmann2004}; \citealt{constantin2008}; \citealt{platenPhD}; \citealt{lopes2017}; \citealt{ceccarelli2021} \citealt{mishra2021}). 

The relatively isolated nature of void galaxies suggests that their growth and evolution do not depend strongly on merger-driven nuclear activity. For example, \cite{ceccarelli2021} found that the growth channels for void galaxies and their central, supermassive black holes differed from those of their non-void counterparts. \cite{ceccarelli2021} attribute this to the fact that, compared to the supermassive black holes in their non-void galaxies, the supermassive black holes in their void galaxies had larger surrounding OIII reservoirs that fed into the central regions of the galaxies. 
%\textcolor{red}{not sure what you're saying here -- is it the ways in which the black holes grow? -- need to clarify what you mean} \textcolor{blue}{OC: This was in reference to how the central blackhole accretes mass over cosmic time. Specifically, the Ceccarelli paper found a higher concentration of AGN in void galaxies in their sample, which they attributed to the fact the SMBH in their void galaxy sample had a larger surrounding OIII reservoir that feeds into the central galaxy.}
In contrast, however, \cite{habouzit} found that void galaxies and their supermassive black holes in the \texttt{Horizon-AGN} simulation grow in a manner 
%\textcolor{red}{?? is it really MANNER or is it RATE ??} 
%\textcolor{orange}{OC Notes: I think manner is more appropriate since the authors were only looking at the $M_{BH}-M_*$ relation of black holes in the z=0 snapshot of Horizon-AGN. Here is a quote from their abstract, though if it helps, "Our results suggest that even if the growth channels in cosmic voids are different from those in denser environments, voids grow their galaxies and BHs in a similar way."} 
that is similar to that of galaxies in denser environments, where merger-driven nuclear activity is common. Understanding the discrepancies between these results could further our understanding of the various ways in which nuclear activity is triggered, as well as the effects that cosmic flow patterns and mergers have on AGN. 

Here we investigate the properties of voids and void galaxies in the $z=0$ snapshot of the \texttt{TNG300} simulation (hereafter TNG300; \citealt{Nelson2018,Springel,Marinacci,Naiman,Pillepich}). 
TNG300 is a cosmological magnetohydrodynamical (MHD) simulation of a $\Lambda$CDM universe with sufficient spatial and mass resolution within a large volume to conduct statistical analyses of voids and void galaxies.  Using TNG300, we construct the largest catalog of luminous void galaxies within a cosmological MHD simulation to date. We compare various physical properties (sizes, colors, star formation rates, luminosity functions, mass functions, metallicities, and nuclear activity) of the void galaxies to those of galaxies found in walls and filaments. In addition, we examine the degree to which location within the cosmic web (i.e., within voids and outside voids) affects AGN activity.

The paper is organized as follows. TNG300 and our void finding algorithm are discussed in \S\ref{sec:methods}. In \S\ref{sec:results} we present the properties of the voids and the void galaxies, and we compare the properties of void galaxies to non-void field galaxies. A summary and discussion of our results is presented in \S\ref{sec:discussion}.  Throughout, we compute error bars using 10,000 bootstrap resamplings of the data.  Error bars are omitted from figures when the error bars are comparable to or smaller than the sizes of the data points.

\section{Methodology}\label{sec:methods}

\subsection{TNG300}\label{sec:simulation}
%The TNG team requests that the simulations be referred to as TNG300, rather than IllustrisTNG300

%If you're going to use h^{-1} in the results section, you need to be consistent and do it here too

We use the $z = 0$ snapshot of the highest-resolution TNG300 simulation to obtain our sample of galaxies and voids. 
TNG300 is the largest simulation box in the IllustrisTNG suite of simulations \citep{Nelson2018,Springel,Marinacci,Naiman,Pillepich}, encompassing a co-moving volume of $205^3 h^{-3} \rm{Mpc}^3$. A total of $2,500^3$ dark matter particles of mass $m_{\rm dm} = 4.0 \times 10^7 h^{-1} M_\odot$ and $2,500^3$ hydrodynamical gas cells with initial baryonic mass $m_{\rm b} = 7.5 \times 10^6 h^{-1} M_\odot$ were used.  In the $z=0$ snapshot, the gravitational force softening is $1.0 h^{-1}$~kpc and the smallest hydrodyamical gas cells are $125 h^{-1}$~pc in extent.  The volume of the simulation is sufficient for the identification of voids with radii as large as $\sim 25 h^{-1}$~Mpc, and the spatial and mass resolution of the simulation is sufficient for the identification of galaxies with stellar masses as small as $10^7 h^{-1} M_\odot$.  
The cosmological parameters adopted in the simulation are: $\Omega_{\Lambda,0} = 0.6911$,
$\Omega_{m,0} = 0.3089$, $\Omega_{b,0} = 0.0486$, $\sigma_8 = 0.8159$, $n_s = 0.9667$, and
$h = 0.6774$ (e.g., \citealt{Planck15}).

To identify luminous galaxies, we use the publicly-available TNG300 subhalo and group catalogs, which list friends-of-friends groups and their substructures. Excluding subhalos that are flagged as being non-cosmological in origin, there are a total of 664,322 luminous galaxies.
%\textcolor{red}{luminous galaxies in total -- this is not the same as your sample, which explicitly excludes cluster galaxies}. \textcolor{blue}{OC: You are right, and I need to double check if that 664,322 number includes cluster galaxies or not. OC 2: I just double checked that this is the total number of luminous galaxies.}
To assign luminosities to the galaxies, we adopt the subhalo magnitudes from the supplementary catalog created by \citet{Nelson2018}. In comparison to the main IllustrisTNG subhalo catalog, the \citet{Nelson2018} catalog better resembles Sloan Digital Sky Survey (SDSS; \citealt{york2000}) photometry because it includes the effects of dust obscuration on the simulated galaxies.

In addition to the group and subhalo catalogs, we also make use of particle data from the $z = 0$ snapshot, including stellar, dark matter, gas, and black hole particles.
Following \cite{Pillepich}, when quoting stellar masses for our TNG300 galaxies, we apply a 40\% correction to the stellar masses from the TNG300 catalog (i.e., to account for the fact that the stellar masses in TNG300 are not as well converged as in the benchmark TNG100 simulation; see Appendix A of \citealt{Pillepich}).

%Here or elsewhere be specific about which parameters are used in what plots/methods

\subsection{Void Finder} \label{sec:voidfinder}
To identify voids, we use the 3D spherical void finder (SVF)  from \cite{padilla2005} as implemented by \cite{paillas2017} and \cite{paillas2019}.\footnote{https://github.com/epaillas/voidfinder/}  This is a computationally inexpensive algorithm that identifies regions with significant central underdensities that converge to the average density beyond the ridges of the voids at $\gtrsim 3$ void radii.
The SVF algorithm can be summarized as follows:

\begin{enumerate}
    \item A rectangular grid is constructed over the galaxy distribution. The number of galaxies within each cell is then counted and any cell that is completely devoid of galaxies (i.e., the cell contains no galaxies) is considered to be a void center.
    \item Spheres are expanded outwards from each void center. The largest sphere around a void center with an integrated underdensity contrast $\Delta_{\rm void}\leq-0.8$ has its radius defined to be the radius of the void. The choice of setting the underdensity contrast threshold to $-0.8$ comes from linear theory arguments in \cite{blumenthal1992}, who showed that voids at the present epoch should have interior densities that are 20\% of the mean density of the {Universe} at the time of shell-crossing.
    \item Any void that neighbors a larger void by more than 20\% of the sum of the radii of both voids is rejected. That is, if the distance, $d$, between voids $i$ and $j$ (where radius $R_i\leq R_j$) satisfies $d \leq 0.2*(R_i + R_j)$, void $i$ is rejected from the sample.
    \item Remaining voids have their centers perturbed in random directions to determine whether their radii can be increased. If a shifted void center results in a larger sphere that satisfies the underdensity contrast criterion from Step 2, the center and radius of the void are then updated to the values of the larger sphere.  
\end{enumerate}

The rectangular grid from Step 1 contains cells that are $5h^{-1}$ Mpc on a side. The grid size was chosen as a balance between: [1] computation time, [2] reducing the error associated with the locations of the centers of the smallest voids, and [3] reducing the number of spurious centers (which can arise due to shot noise). That is, if a void center is identified with low resolution, the error associated with the radius of that void will be adversely affected (i.e., since the radius depends on the integrated underdensity contrast threshold in the region surrounding that center). Therefore, a cell size is chosen such that the error it contributes to the radius of the smallest voids in our sample is less than 10\% of the total error for that result (see \citealt{paillas2017} for a detailed discussion).

The 3D SVF does not provide the detailed description of void geometry that can be obtained with more sophisticated void finding algorithms (see, e.g., \citealt{neyrinck2008}; \citealt{platen2007}); however, voids in the local {Universe} tend to exhibit spherical symmetry (see, e.g., \citealt{Icke}; \citealt{sheth}; \citealt{sutter2014}; \citealt{hamaus2016}).  
Moreover, \cite{paillas2019} compared the results of six void finding techniques in the context of differentiating $f(R)$ gravity models and found that similar results were obtained regardless of the choice of void finding algorithm. For our purposes, then, we consider the 3D SVF to be a reasonable choice of algorithm.

\subsection{Field Galaxy Sample} \label{sec:galselection}

Our investigation of galaxy properties focuses on field galaxies (i.e., galaxies that reside outside cluster environments), subdivided according to whether the galaxies reside within a void (``void galaxies'') or outside a void (``non-void'' galaxies).  We define field galaxies to be all galaxies that are not contained within parent halos with masses $ \ge 10^{14} h^{-1} M_\odot$, which eliminates cluster galaxies from the sample.  We then use the void catalog from \S\ref{sec:voidfinder} to separate void galaxies from non-void galaxies. From this, our field galaxy sample consists of 75,220 void galaxies and 527,454 non-void galaxies.

As part of our analysis, we investigate the relative ages of void and non-void galaxies. Even in simulation space, determining the formation time of a galaxy is not simple.  This is due to the hierarchical nature of structure formation, which results in many smaller overdensities merging to form a single galaxy. Here we adopt two indicators of galaxy age: [1] the age of the oldest bound stellar particle and [2] the luminosity-weighted age. For both of these age indicators, we use the subhalo cutout provided by the TNG, which contains all particles that are identified by the \texttt{SUBFIND} algorithm as being bound to the halo. Formation times of stellar particles are given as scale factors in the simulation, and we convert these to lookback times using the \cite{Planck15} cosmological parameters and the \texttt{astropy.cosmology} package \citep{astropy:2013,astropy:2018}. 

The age of the oldest bound stellar particle is a useful indicator of the age of a simulated galaxy, but it is not straightforward to use as a comparison to observed galaxies.  In contrast, the luminosity-weighted age is a metric that can be compared to observations (see, e.g., \citealt{Li18}).  To obtain luminosity-weighted ages of the simulated galaxies we use the following equation from \cite{Lu20}:
\begin{equation} \label{eq:L_weight}
    \langle x \rangle = \frac{\sum_{k=1}^{N}L_k x_k}{\sum_{k=1}^{N} L_k},
\end{equation}
where $x_k$ is the logarithm (base 10) of the formation age of the $k^{\rm th}$ particle and $L_k$ is the SDSS $r$-band luminosity of the $k^{\rm th}$ particle. Above, the summation is computed over all stellar particles that are bound to the subhalo.

\section{Results}\label{sec:results}
\subsection{Void Properties}

\begin{figure}
    \centering
    \includegraphics[width=0.75\columnwidth]{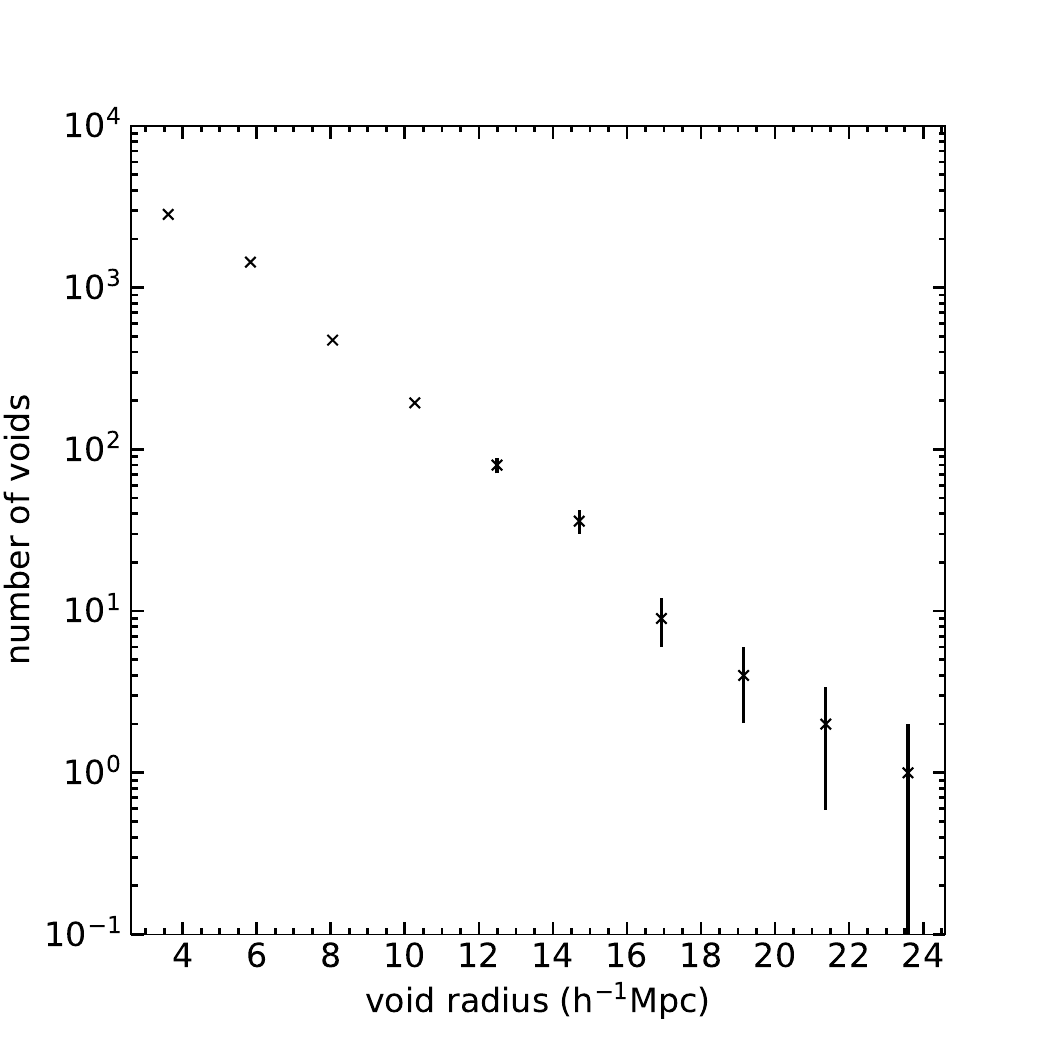}
    \caption{Distribution of void radii.}
    \label{fig:void_radii}
\end{figure}

Using the 3D SVF algorithm, a total of 5,078 voids were identified.  The void radii span an order of magnitude, with the smallest voids having radii of $r_v = 2.5 h^{-1}$~Mpc and the largest void having a radius of $r_v = 24.7 h^{-1}$~Mpc.  The median void radius is $4.4 h^{-1}$~Mpc.  In total, voids cover a volume of $168.3^3h^{-3}$~Mpc$^3$, or 82\% of the simulation box.  The number of voids as a function of radius is shown in Figure \ref{fig:void_radii}, from which it is clear that the majority of voids have radii $\lesssim 10 h^{-1}$~Mpc.  

\begin{figure}
    \centering
    \includegraphics[width=0.95\columnwidth]{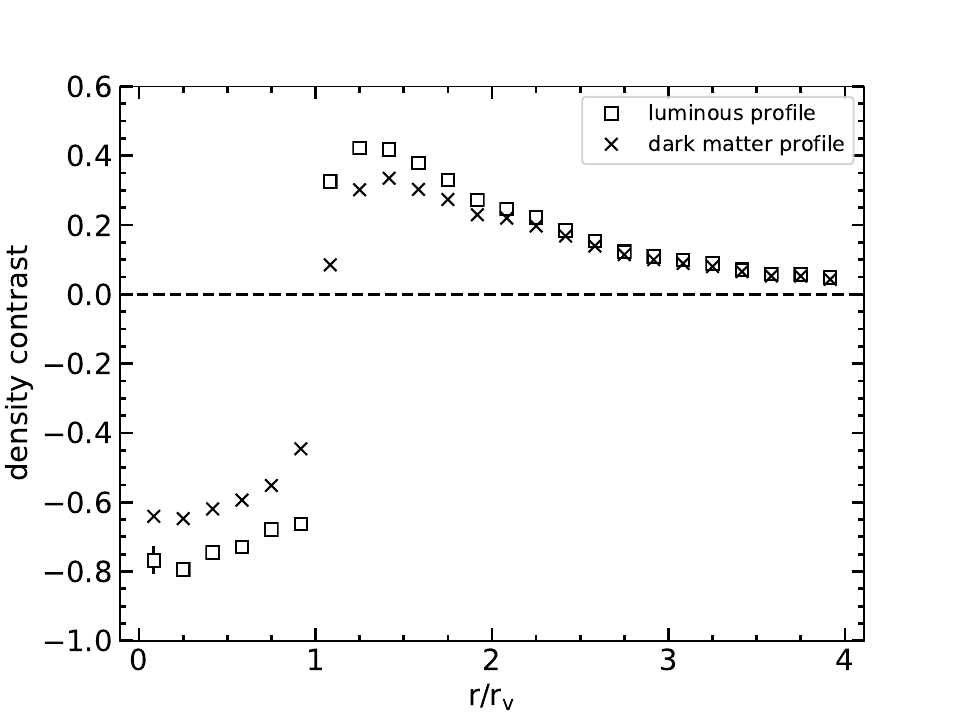}
    \caption{Average radial number density contrast profiles for the voids, computed using concentric spherical shells centered on the void centers.  Radii of the shells are given in units of the void radius, $r_v$. Squares: number density contrast of the luminous galaxies.  Crosses: number density contrast of the dark matter particles.}
    \label{fig:avg_rad_profiles}
\end{figure}

The mean radial number density contrast of the voids is shown in Figure \ref{fig:avg_rad_profiles}.  Squares show the result obtained using the luminous galaxies (``luminous density contrast'') and crosses show the result obtained using the dark matter particles (``dark matter density contrast''). For each void, the density contrast in concentric spherical shells of thickness $\delta r$, centered on the void center, was calculated as

\begin{equation}
    \Delta = \frac{n(r)}{\bar{n}(r)} - 1,
\end{equation}

\noindent 
where $n(r)$ is the number of galaxies or dark matter particles within the radial bin that ranges from $r-\frac{\delta r}{2}$ to $r+\frac{\delta r}{2}$ and $\bar{n}(r)$ is the average number density of galaxies or dark matter particles within the simulation box. 
In both cases (i.e., luminous density contrast and dark matter density contrast), the mean radial density profiles resemble reverse spherical top-hat distributions, which are expected for underdense regions in the local {Universe} (see, e.g., \citealt{Icke}; \citealt{sheth}). These profiles are characterized by an underdense, flat interior that rises steeply above the mean density at the ridge of a void and slowly decreases to the mean density at large distances from the void center.

From Figure \ref{fig:avg_rad_profiles}, it is clear that mass does not directly trace light within the voids.  Rather, the mean central density contrast obtained from the galaxies ($\Delta = -0.77 \pm 0.04)$ is somewhat lower than the mean central density contrast of the dark matter ($\Delta = -0.64 \pm 0.02$); i.e., the centers of the voids are more ``devoid'' of galaxies than they are of dark matter.  The opposite is true at the ridges of the voids where, on average, there is a higher concentration of galaxies than dark matter, with the mean dark matter density contrast reaching a maximum of $0.33 \pm 0.02$ at $r=1.25r_v$ and the mean luminous density contrast reaching a maximum contrast of $0.42 \pm 0.02$ at $r=1.4r_v$.

\begin{figure}
    \centering
    \includegraphics[width=0.95\columnwidth]{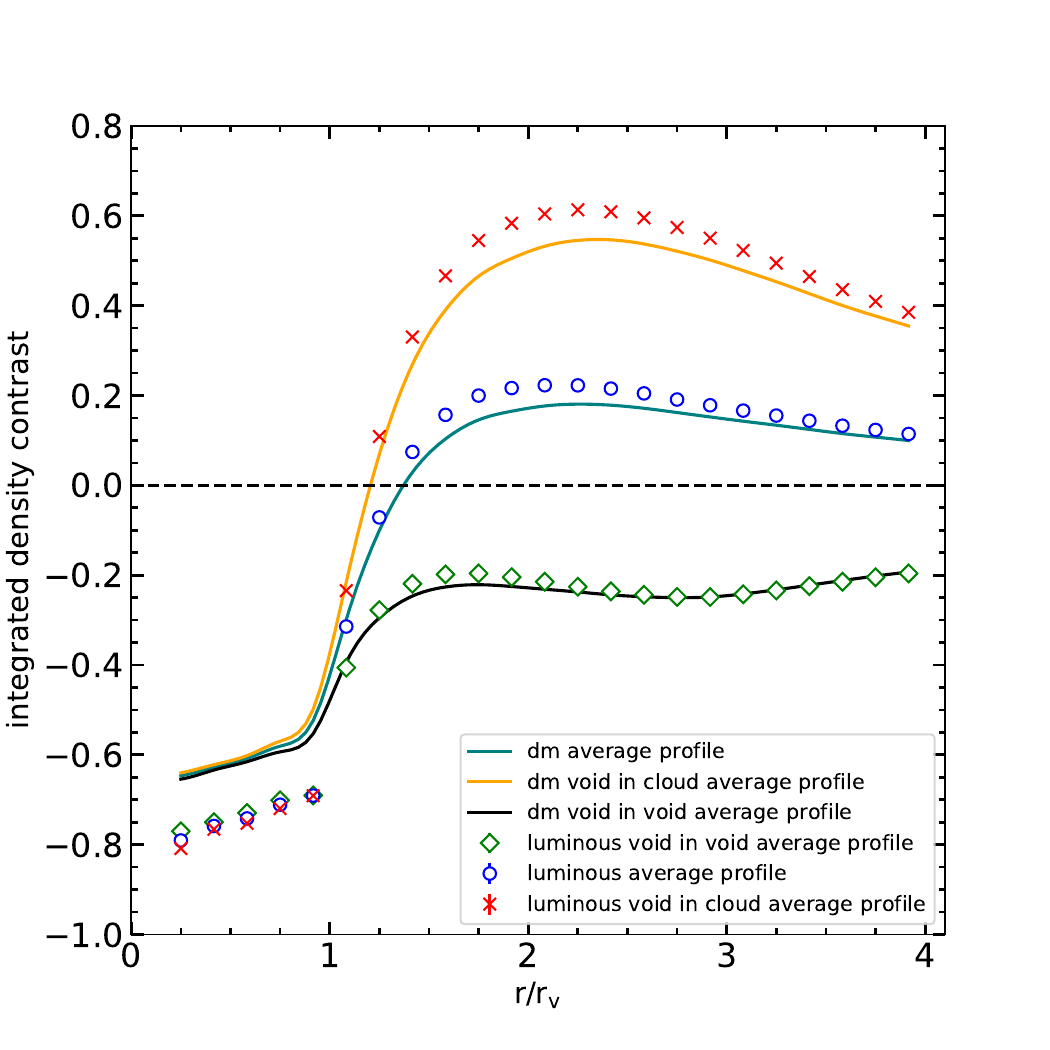}
    \caption{Mean integrated number density contrast computed using luminous galaxies (points) and dark matter particles (lines). Blue line and blue circles: results obtained using all voids in the simulation. Yellow line and red crosses: results for voids in regions of the simulation that have a high local background density (``voids-in-clouds'').  Black line and green diamonds: results for voids in regions of the simulation that have a low local background density (``voids-in-voids'').  See text for definitions of voids-in-clouds and voids-in-voids.}
    \label{fig:avg_int_profiles}
\end{figure}

Voids form in regions of space with differing background densities (i.e., as compared to the average density of the {Universe}).  That is, some voids form in regions of space that have relatively high local background densities (``voids in clouds'') while others form in regions of space that have relatively low local background densities (``voids in voids''); see \cite{sheth} for a discussion of void hierarchy.  In Figure~\ref{fig:avg_int_profiles} we use mean integrated number density profiles to explore the effects of the local background density on the density contrast.  For legibility of the figure, the number density profiles for the dark matter particles are shown as spline fits. Following \cite{sheth},
we define voids-in-voids to be those with an integrated galaxy density contrast $< 0$ at $r=3r_v$ and voids-in-clouds to be those with an integrated galaxy density contrast $> 0$ at $r=3r_v$.
To construct the mean integrated density profiles, the number density contrast of the galaxies (points) and the dark matter particles (lines) was computed in concentric spheres of radius $r$, centered on each void center, from which means were then computed.

The mean integrated number density profiles obtained using all voids (blue line and blue crosses in Figure~\ref{fig:avg_int_profiles}) show the same trends as the mean differential number density profiles in Figure~\ref{fig:avg_rad_profiles}.  On average, the central regions of the voids have a dark matter density contrast that is $65\% \pm 2\%$ lower than the average dark matter density contrast in the simulation and a galaxy density contrast that is $77\% \pm 4\%$ lower than the average galaxy density contrast in the simulation.  At the ridges of the voids, the mean interior density contrast for the galaxies reaches a maximum that is $22\% \pm 1\%$ higher than the average galaxy density contrast at $r = 2.08r_v$ and the mean interior density contrast for the dark matter reaches a maximum that is $18\% \pm 2\%$ higher than the average dark matter density contrast at $r = 2.25 r_v$.

For radii $r \lesssim  r_v$, the local background density in which the voids are located has relatively little effect on the integrated number density profiles of the galaxies and the dark matter.  For radii $r > r_v$, however, significant differences for the integrated number density profiles of voids-in-voids (black line and green diamonds) and voids-in-clouds (yellow line and red crosses) occur. In the case of voids-in-voids, galaxies largely trace the dark matter for $r \gtrsim r_v$.  At the ridges of voids-in-voids, the mean interior density contrast remains less than the average in the simulation (i.e., the mean interior density contrast is $-0.20\pm 0.01$ at the ridges) and for radii $1.5 r_v \lesssim r \lesssim 4r_v$ the mean interior density contrast of voids-in-voids remains approximately constant at a value of $\sim -0.2$.  
In the case of voids-in-clouds, the galaxies trace the dark matter for a 
small range of radii ($r_v \lesssim r \lesssim 1.2 r_v$), but are more overdense than the dark matter for $r \gtrsim 1.2r_v$.  At the ridges of voids-in-clouds, the integrated number densities of the galaxies and dark matter exceed the integrated number densities of the galaxies and dark matter in the full void sample by a factor of $\sim 3$ and they remain significantly higher than than the average density out to distances as large as $\sim 4 r_v$.

\subsection{Properties of Void and Non-void Galaxies}\label{sec:voidgalaxies}

Below we investigate the following properties of void and non-void galaxies: [1] distribution of physical sizes, [2] optical color-magnitude relationships, [3] luminosity and stellar mass functions, [4] ages, [5] stellar and gas chemical abundances, [6] specific star formation rates, and [7] relationships between stellar mass and supermassive black hole mass, as well as AGN fraction.

\subsubsection{Physical Sizes}

\begin{figure}
    \centering
    \includegraphics[width=0.75\columnwidth]{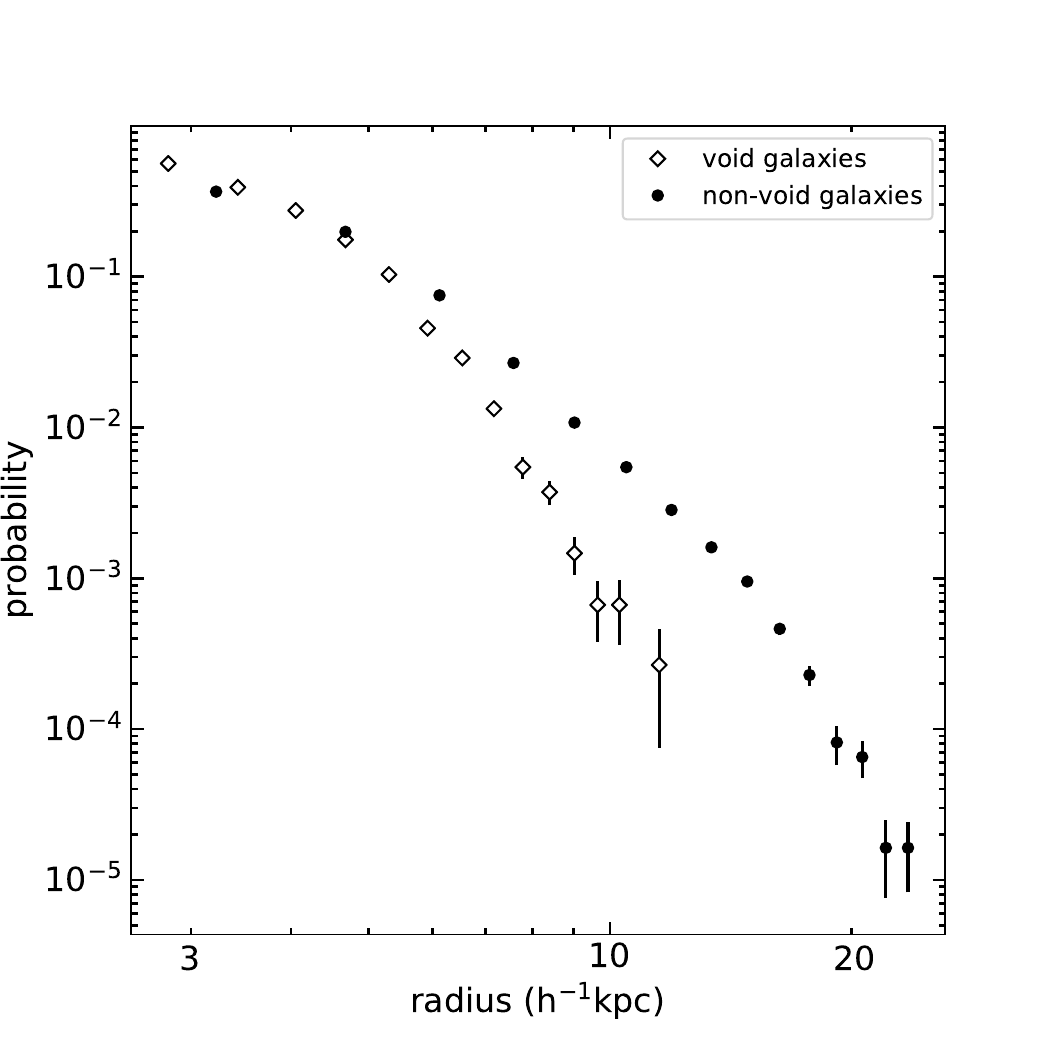}
    \caption{Probability distributions for the radii of void galaxies (diamonds) and non-void galaxies (circles).}
    \label{fig:galaxy_radii}
\end{figure}

The TNG300 subhalo catalog defines the photometric radii of the galaxies to be the radii at which the surface brightness profiles drop below 20.7~mag~arcsec$^{-1}$ in the $K$-band.  We use this definition to compute normalized probability distributions for the radii of void and non-void galaxies, results of which are shown in 
Figure \ref{fig:galaxy_radii}.  From this figure, the majority of the 
galaxies in both populations have radii $\lesssim 5h^{-1}$~kpc, and the probability of a given galaxy having a radius $\lesssim 5h^{-1}$~kpc is the same for both void and non-void galaxies. For galaxy radii $\gtrsim 5h^{-1}$~kpc, the distribution of void galaxy radii declines much more steeply than the distribution of non-void galaxy radii.  As a result, the largest void galaxies have radii $\sim 11h^{-1}$~kpc, while the largest non-void galaxies have radii that are $\sim 2$ times larger than the largest void galaxies.

\subsubsection{Optical Color-Magnitude Relationships}

\begin{figure}
    \centering
    \includegraphics[width=0.95\columnwidth]{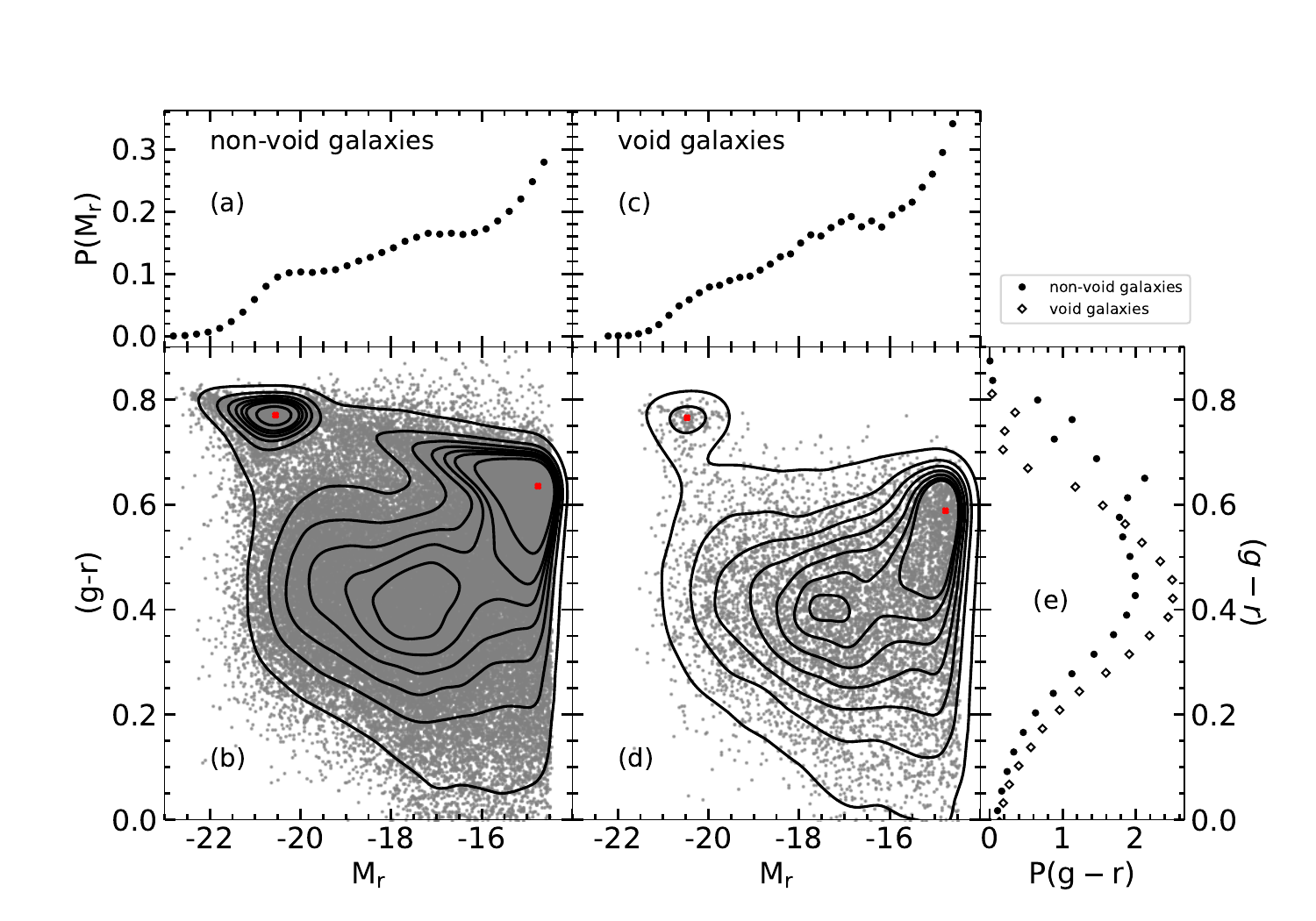}
    \caption{Optical color-magnitude relations for non-void galaxies (panel b) and void galaxies (panel d). Black contours: linearly spaced density contours, computed using all galaxies in each sample. Gray points: 10\% of the non-void galaxies, randomly selected from the complete sample (panel b) and 100\% of the void galaxies (panel d). Red crosses indicate the red and blue peaks for each of the distributions. Top panels: normalized probability distributions for the $r$-band absolute magnitudes of non-void galaxies (panel a) and void galaxies (panel c). Side panel: normalized probability distributions of the $(g-r)$ colors of void galaxies (diamonds) and non-void galaxies (circles).}
    \label{fig:gr_vs_r}
\end{figure}

Figure \ref{fig:gr_vs_r} shows the relationship between the $(g-r)$ optical color and the absolute SDSS $r$-band magnitude, $M_r$, for non-void galaxies (panel b) and void galaxies (panel d). From these, it is clear that both void and non-void galaxies have distributions that peak in two locations: [1] intrinsically bright, red galaxies and [2] intrinsically faint, blue galaxies.  Red crosses in panels (b) and (d) of Figure \ref{fig:gr_vs_r} indicate the locations of the peaks, and show that the peaks occur in similar locations of the $(g-r)$ vs.\ $M_r$ space for both void and non-void galaxies.
Formally, the blue peak occurs at $(M_r, g-r) = (-14.77, 0.59)$ for the void galaxies and at $(M_r, g-r) = (-14.76, 0.64)$ for the non-void galaxies; i.e., the blue peaks occur at identical absolute magnitudes but slightly different colors, with the void galaxies being somewhat bluer than the non-void galaxies near the blue peaks. The red peak occurs at $(M_r, g-r) = (-20.47, 0.77)$ for the void galaxies and at $(M_r, g-r) = (-20.55, 0.77)$ for the non-void galaxies; i.e., the red peaks occur at essentially identical absolute magnitudes and colors for both populations of galaxies. 

Panels (a) and (c) in Figure \ref{fig:gr_vs_r} show normalized probability distributions for the absolute $r$-band magnitudes of the non-void and void galaxies, respectively.  From this, it is clear that distributions differ signficantly, and this will be reflected in the luminosity functions of void and non-void galaxies below.  Panel (e) in Figure~\ref{fig:gr_vs_r} shows normalized probability distributions for the $(g-r)$ colors of void galaxies (diamonds) and non-void galaxies (circles).  From this, it is clear that, while both types of galaxies have a broad distribution of optical colors, the distributions are substantially different and, in particular, there is a much higher concentration of red, non-void galaxies than red void galaxies.  
%\textcolor{red}{Why is the red end of the distribution seem so noisy for the non-void galaxies?  Is it because it's actually tri-modal, not bi-modal?  If so, this could be revealed by doubling the number of points used for the non-void galaxies.} \textcolor{blue}{OC: I've doubled the number of points used displayed in the non-void galaxies panel. I think there are more galaxies in the 'green valley' of the non-void galaxy distribution that are contributing to the noisiness in the red-end of panel (e).}

\begin{figure}
    \centering
    \includegraphics[width=0.65\columnwidth]{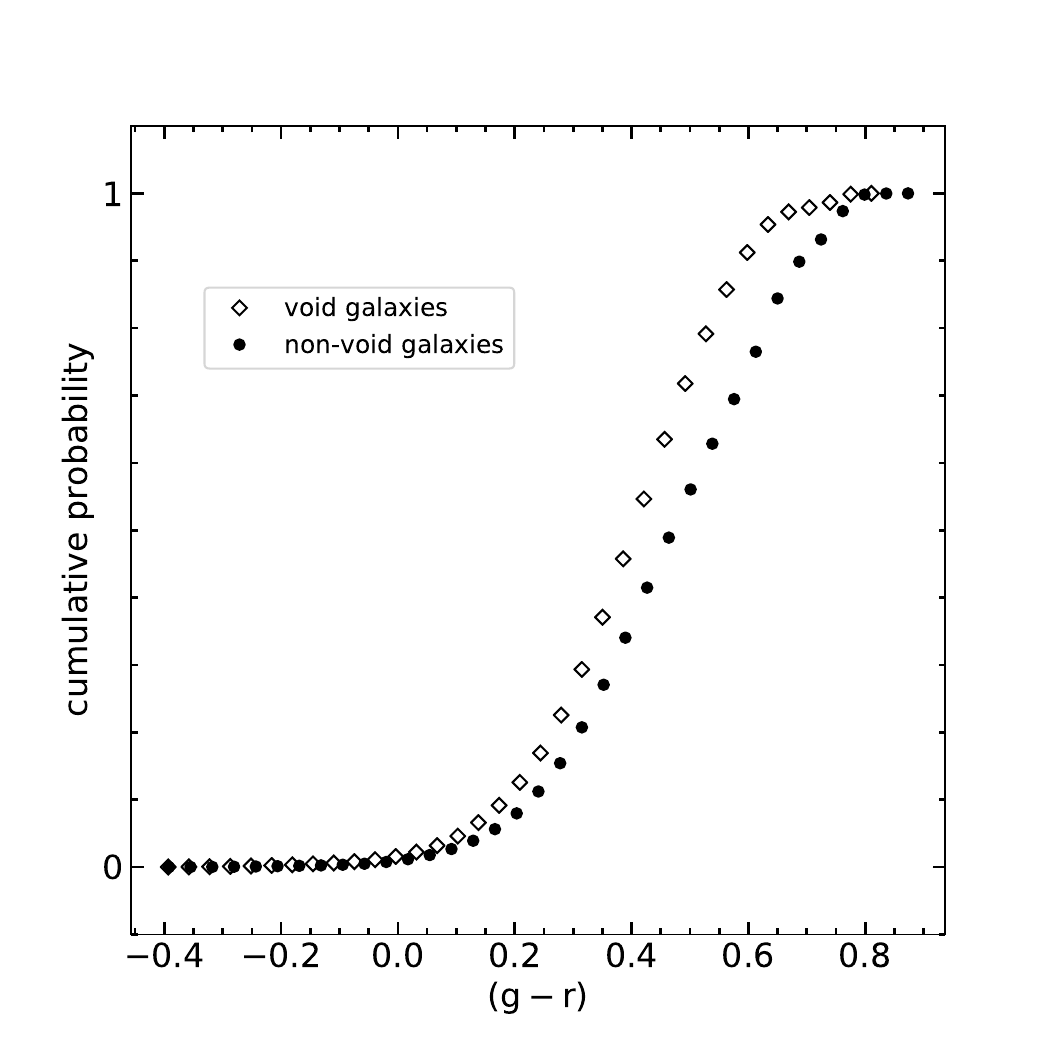}
    \caption{Cumulative probability distribution functions for $(g-r)$ galaxy colors. Diamonds: void galaxies.  Circles: non-void galaxies.}
    \label{fig:cumulative_colors}
\end{figure}

We further explore the differences between the optical color distributions for void and non-void galaxies in Figure \ref{fig:cumulative_colors}, which shows the cumulative probability distributions of $(g-r)$ values for void galaxies (diamonds) and non-void galaxies (circles).  A two-sample Kolmogorov-Smirnov (KS) test performed on the two distributions in Figure \ref{fig:cumulative_colors} rejects the null hypothesis that both distributions are drawn from the same underlying distribution at a high confidence level ($> 99.9999$\%), indicating that the distribution of void galaxy colors is significantly different from that of non-void galaxies.  In particular, void galaxies are typically bluer than non-void galaxies, with the median $(g-r)$ color of void galaxies being $0.4205 \pm 0.0007$ and the median $(g-r)$ color of non-void galaxies being $0.4881 \pm 0.0003$.

\subsubsection{Luminosity and Stellar Mass Functions}

\begin{figure}
    \centering
    \includegraphics[width=0.75\columnwidth]{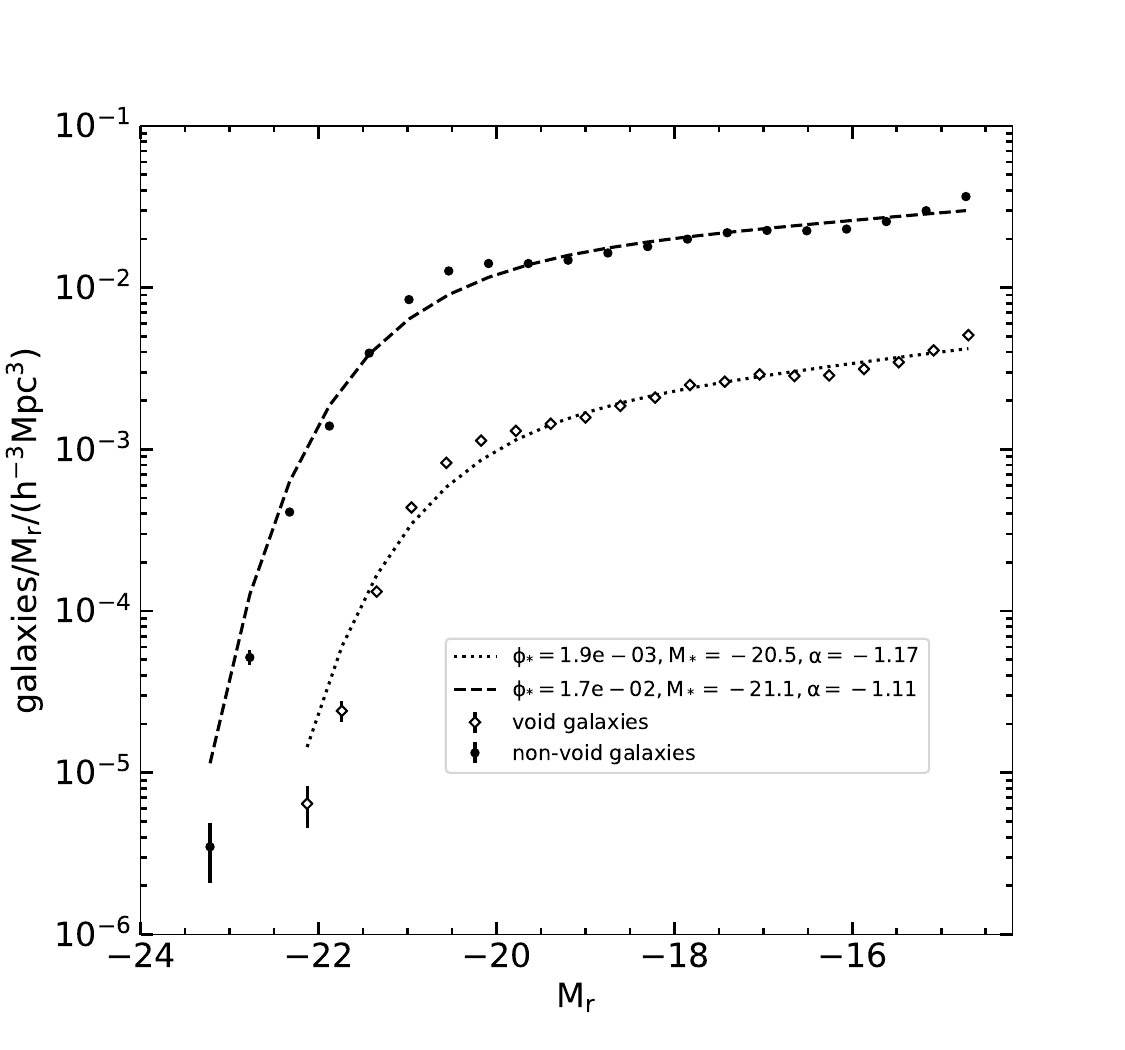}
        \caption{Luminosity functions in the SDSS $r$-band for void galaxies (diamonds) and non-void galaxies (circles). Dotted line: best fitting Schechter luminosity function for void galaxies. Dashed line: best fitting Schechter luminosity function for non-void galaxies.}
    \label{fig:lum_func_r}
\end{figure}

Next we compute luminosity functions and stellar mass functions for void and non-void galaxies.  Figure~\ref{fig:lum_func_r} shows the luminosity functions, computed in terms of SDSS $r$-band absolute magnitude (i.e., the number of galaxies per magnitude per unit volume). Lines in Figure \ref{fig:lum_func_r} show the best-fitting Schechter luminosity functions \citep{schechter1976} of the form:

\begin{equation}
    \Phi(M) dM = 0.4 \ln(10)~ \Phi_\ast ~\left[ 10^{0.4 (M_\ast - M) }\right]^{(1+\alpha)} \exp\left[-10^{0.4 (M_\ast - M)} \right] ~ dM ~~~~~~~,
\end{equation} 

\noindent
where $M_\ast$ is the absolute magnitude of an $L_\ast$ galaxy.

Overall, the luminosity functions of both void and non-void galaxies are fitted reasonably well by Schechter functions, though some deviation is apparent (particularly at the extreme ends of the luminosity functions).  The parameters of the best-fitting Schechter functions are $\phi_\ast=(1.9\pm 0.2)\times 10^{-3}$~Mpc$^{-3}$, $M_\ast = -20.5\pm 0.1$, $\alpha=-1.17\pm 0.03$ (non-void galaxies) and $\phi_\ast =(1.7\pm 0.2)\times 10^{-3}$~Mpc$^{-3}$, $M_\ast =-21.1\pm 0.1$, $\alpha=-1.11\pm 0.02$ (void galaxies). Within the formal errors, the luminosity functions have characteristic bright galaxy absolute magnitudes, $M_\ast$, that differ significantly; i.e., $L_{\ast,r}$ non-void galaxies are intrinsically $\sim 70$\% brighter than $L_{\ast,r}$ void galaxies.  However, the faint end slopes of the luminosity functions, $\alpha$, are consistent within $2\sigma$.

\begin{figure*}
    \centering
    \includegraphics[width=0.95\textwidth]{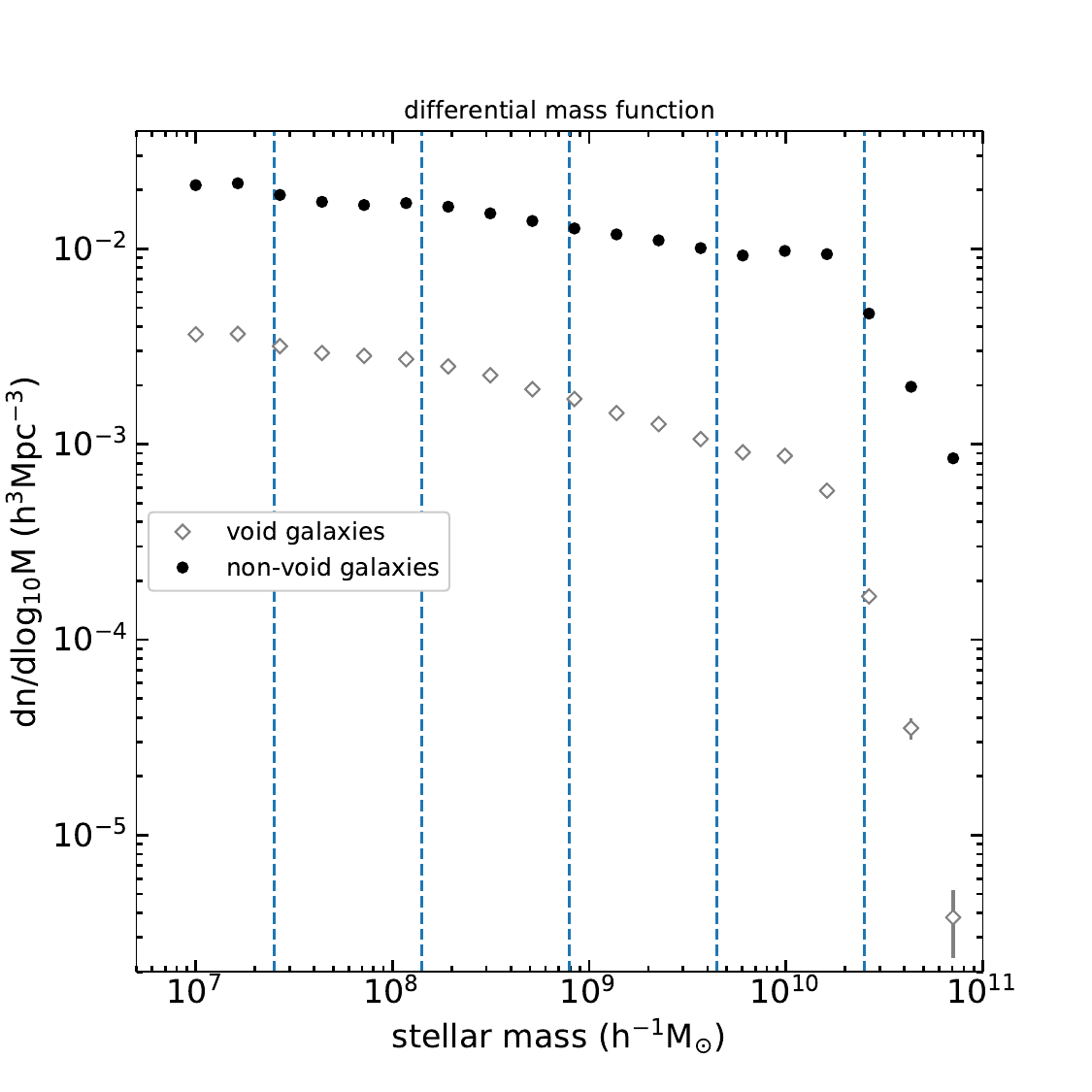}
    \caption{Differential stellar mass functions for void galaxies (diamonds) and non-void galaxies (circles). Dashed lines indicate the boundaries of the stellar mass bins used in the analysis of stellar and gas metallicities (see text).}
    \label{fig:massfuncs}
\end{figure*}

Results for the differential stellar mass functions (i.e., the number of galaxies per log stellar mass bin per unit volume) of void and non-void galaxies are shown in Figure~\ref{fig:massfuncs}. For stellar masses between $10^{7.25} h^{-1} M_\odot$ and $10^{10.25} h^{-1} M_\odot$, the mass functions exhibit roughly power law behaviors, with the slope of the power law being substantially shallower for non-void galaxies than it is for void galaxies. Below stellar masses of $10^{7.25} h^{-1} M_\odot$, both mass functions fall below the power law due to the finite resolution of the simulation, which leads to undercounting of the smallest galaxies.  Above stellar masses of $10^{10.25} h^{-1} M_\odot$, both mass functions decrease sharply due to the fact that high mass galaxies are relatively rare objects.  From the high mass ends of the mass functions, however, it is clear that the sample of non-void galaxies has a larger fraction of high mass galaxies than does the void galaxy sample.  The combined differences in the mass functions result in the median stellar mass for the void galaxies being a factor of $\sim 2$ smaller than the median stellar mass for the non-void galaxies: $(1.58 \pm 0.01)\times 10^8 h^{-1} M_\odot$ vs.\ $(3.08 \pm 0.01)\times 10^8 h^{-1} M_\odot$. 
%$(1.13 \pm 0.01)\times 10^8 h^{-1} M_\odot$ vs.\ $(2.20 \pm 0.01)\times 10^8 h^{-1} M_\odot$.
%median stellar mass for the void galaxies is $10^{8.054\pm 0.005}h^{-1}$ $M_\odot$, while the median stellar mass for the non-void galaxies is $10^{8.342\pm0.002}h^{-1}$ $M_\odot$}.

%\textcolor{red}{In what way does the plot of the cumulative mass function add value here?  It doesn't seem that you've discussed it.} \textcolor{blue}{OC: Honestly, I think the differential mass function is sufficient for my analysis. I was hesistant on removing the cumulative function in case some readers prefer to look at those over the differential function, but I would not be opposed to removing that plot.} \textcolor{blue}{OC-2: I've removed the cumulative mass function from figure 8, and I believe I've removed all reference to it from the text.}

\subsubsection{Ages}

Next, we quantify the ages of the void and non-void galaxies using two metrics: [1] the oldest stellar particles within the subhalos and [2] the luminosity weighted ages of the galaxies (see Equation~\ref{eq:L_weight}). From this, the median age of the oldest stellar particles bound to the subhalos in the void galaxies is 10.1033 $\pm$ 0.0001~Gyr and the median age of the oldest stellar particles bound to the subhalos in the non-void galaxies is 10.10685 $\pm$ 0.00003~Gyr; i.e., the oldest stellar particles bound to the subhalos in the void galaxies are $\sim 3.6$~Myr years younger than the oldest stellar particles in the non-void galaxies.  In comparison, the median luminosity weighted age of the void galaxies is 9.217 $\pm$ 0.002~Gyr and the median luminosity weighted age of the non-void galaxies is 9.3642 $\pm$ 0.0009~Gyr; i.e., when using luminosity weighted ages, we find that the void galaxies are $\sim 147$~Myr younger than the non-void galaxies.  Both age indicators reveal that void galaxies are systematically younger than non-void galaxies, but the age difference between void and non-void galaxies is reflected less by the time at which the first stars formed (i.e., the ages of the oldest stellar particles) than it is by the overall star formation histories of the galaxies (i.e., the luminosity weighted ages). {That being said, the difference between the median luminosity-weighted ages of void and non-void galaxies is too small to be detected in modern surveys which report typical dispersions for luminosity-weighted ages of galaxy populations between $0.15-0.30$ dex (see, e.g., \citealt{gonzalez2014}; \citealt{scott2017}; \citealt{Li18}; \citealt{Lu20})}.%. For instance, \cite{gonzalez2014} report a dispersion of luminosity-weighted ages of 0.1 dex for galaxies in The Calar Alto Integral Field Area \citep{califa1, CALIFA2} survey, \cite{scott2017} report a typical uncertainty of 0.15 dex for the ages of galaxies in the Sydney-AAO Multi-object Integral field spectrograph galaxy survey \citep{SAMI}, and \cite{Lu20} report error bars on their luminosity-weighted ages of $0.35-0.15$ dex for galaxies with velocity dispersions between $1.7-2.4$ $\rm km \: s^{-1}$ in the Mapping Nearby Galaxies at Apache Point Observatory survey \citep{manga}.}

%results of the luminosity weighted ages suggest void galaxies are $\sim$ 140 million years younger than their non-void counterparts. \textcolor{blue}{OC: Is this an accurate statement to make?} Yes -- though it's closer to 147 million years.  The error bar for the age difference between the oldest stellar particles is 0.1 Myr; the error bar for the age difference between the luminosity weighted ages is 2 Myr.

\subsubsection{Stellar and Gas Chemical Abundances}

\begin{figure*}
    \centering
    \includegraphics[width=0.95\textwidth]{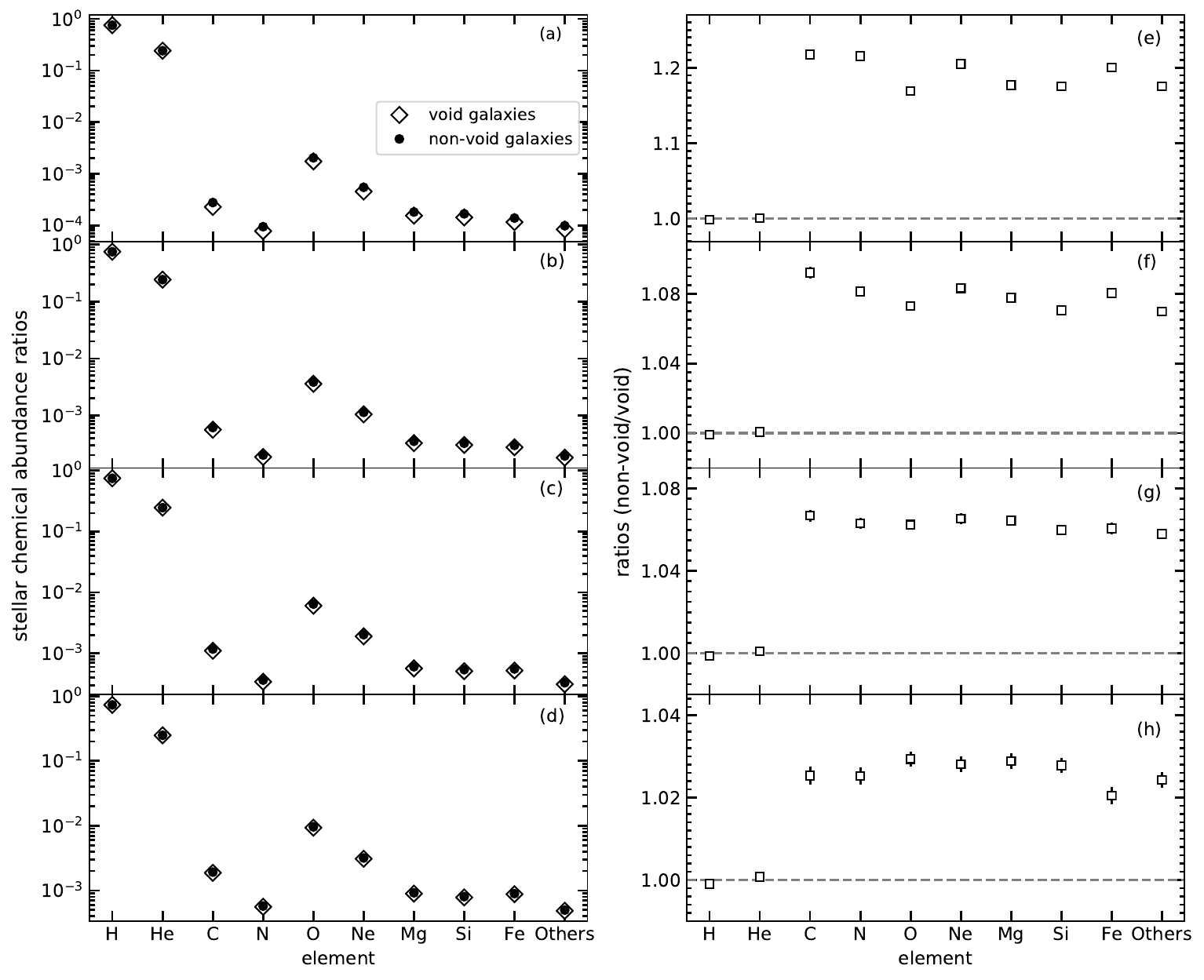}
    \caption{{\it Left:} Average stellar chemical abundance ratios for void galaxies (diamonds) and non-void galaxies (circles). {\it Right:} Ratios of non-void and void galaxy stellar chemical abundances from the left panels. Galaxies with stellar masses $10^{7.40} h^{-1} M_\odot \le M_* < 10^{8.15}h^{-1}$ $ M_\odot$: panels (a) and (e). Galaxies with stellar masses $10^{8.15} h^{-1} M_\odot \le M_* < 10^{8.90}h^{-1}$ $ M_\odot$: panels (b) and (f). Galaxies with stellar masses $10^{8.90} h^{-1} M_\odot \le M_* < 10^{9.65}h^{-1}$ $ M_\odot$: panels (c) and (g). Galaxies with stellar masses $10^{9.65} h^{-1} M_\odot \le M_* < 10^{10.40}h^{-1}$ $ M_\odot$: panels (d) and (h). Note that, for clarity, different ordinate ranges are used in the right panels.}
    \label{fig:stellar_metallicities}
\end{figure*}

\begin{figure*}
    \centering
    \includegraphics[width=0.95\textwidth]{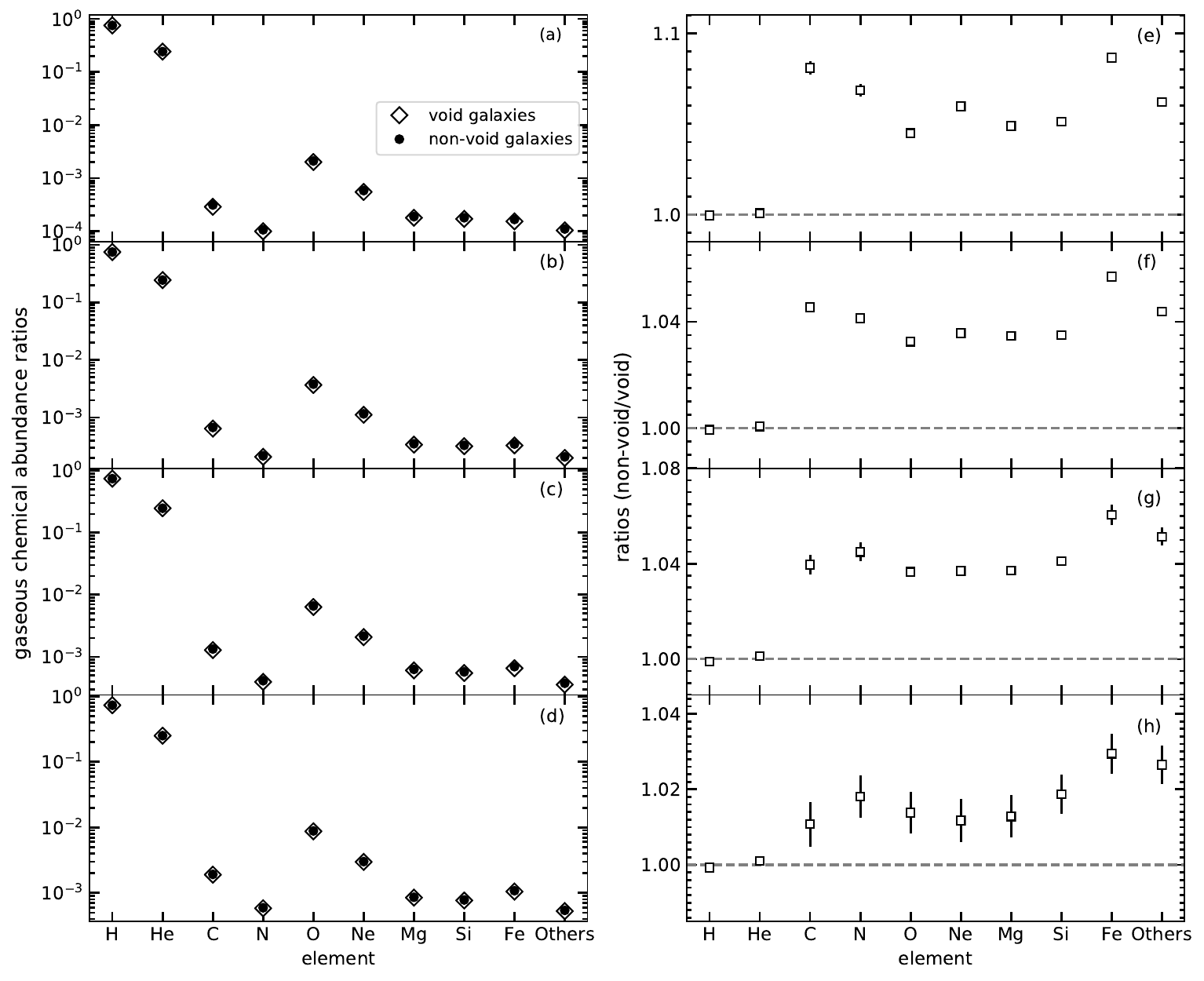}
    \caption{Same as Figure~\ref{fig:stellar_metallicities}, but for chemical abundance ratios of the gas.}
    \label{fig:gas_metallicities}
\end{figure*}

Results for the stellar and gas chemical abundance ratios are shown in Figures~\ref{fig:stellar_metallicities} and \ref{fig:gas_metallicities}, respectively. Here we subdivide the galaxy samples using four distinct stellar mass bins, the boundaries of which 
are $M_*=10^{7.40} h^{-1} M_\odot$, $ 10^{8.15} h^{-1} M_\odot$, $ 10^{8.90} h^{-1} M_\odot$, $ 10^{9.65} h^{-1} M_\odot$, and $ 10^{10.40}h^{-1} M_\odot$. These are indicated by vertical, dashed blue lines in Figure~\ref{fig:massfuncs}. The panels in Figures~\ref{fig:stellar_metallicities} and \ref{fig:gas_metallicities} are arranged vertically in order of increasing stellar mass, such that panels (a) and (e) include only galaxies in the first bin, panels(b) and (f) include only galaxies in the second bin, panels (c) and (g) include only galaxies in the third bin, and panels (d) and (h) include only galaxies in the fourth bin.
The left panels of Figures \ref{fig:stellar_metallicities} and \ref{fig:gas_metallicities} show the average stellar and gas chemical abundance ratios, respectively.  The right panels of Figures \ref{fig:stellar_metallicities} and \ref{fig:gas_metallicities} show the ratios of the corresponding data points for void and non-void galaxies from the left panels.

From Figures~\ref{fig:stellar_metallicities} and \ref{fig:gas_metallicities}, void galaxies have stellar and gas metal fractions that are systematically lower than those of non-void galaxies. This result holds true across all stellar mass bins, but the differences are more pronounced for galaxies with lower stellar masses than they are for galaxies with higher stellar masses. That is, in order of increasing stellar mass, the stellar metallicities of non-void galaxies exceed those of void galaxies by $\sim 20$\% (Figure~\ref{fig:stellar_metallicities}e), $\sim 8$\% (Figure~\ref{fig:stellar_metallicities}f), $\sim 6.5$\% (Figure~\ref{fig:stellar_metallicities}g), and $\sim 2.5$\% (Figure~\ref{fig:stellar_metallicities}h).  A similar, but less pronounced, trend is shown by the gas metallicities; i.e., in order of increasing stellar mass, the gas metallicities of non-void galaxies exceed those of void galaxies by $\sim 7.5$\% (Figure~\ref{fig:gas_metallicities}e), $\sim 4.5$\% (Figure~\ref{fig:gas_metallicities}f and g), and $\sim 2$\% (Figure~\ref{fig:gas_metallicities}h).

\subsubsection{Specific Star Formation Rates}

\begin{figure*}
    \centering
    \includegraphics[width=0.95\columnwidth]{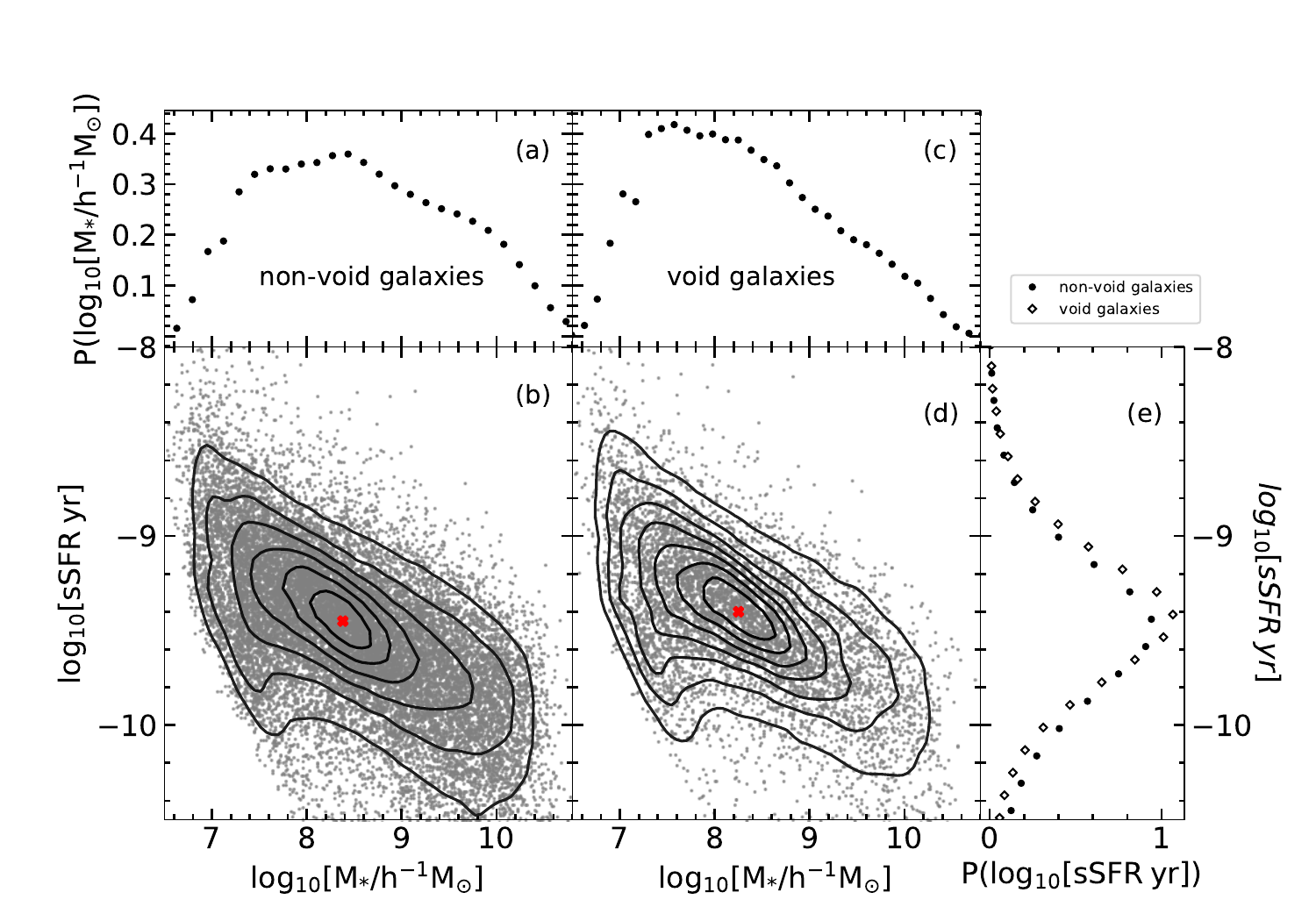}
    \caption{Specific star formation rate vs.\ stellar mass for non-void galaxies (panel b) and void galaxies (panel c). Black contours: density contours, evenly spaced in the logarithm, computed using all galaxies in each sample. Gray points: 5\% of the non-void galaxies, randomly selected from the complete sample (panel b) and 100\% of the void galaxies (panel d). Red crosses indicate the central peaks of the distributions. Top: Normalized 1D probability distributions for the stellar masses of non-void galaxies (panel a) and non-void galaxies (panel b).  Side: Normalized 1D probability distributions for the sSFR of void galaxies (diamonds) and non-void galaxies (circles).}
    \label{fig:ssfr}
\end{figure*}

Results for the relationships between stellar mass and specific star formation rate (sSFR) are shown in Figure~\ref{fig:ssfr}b) (non-void galaxies) and Figure~\ref{fig:ssfr}c) (void galaxies).  Here, the specific star formation rates are instantaneous rates, derived from the sum of star formation rates in individual gas cells at $z=0$. From Figure~\ref{fig:ssfr}b) and c), it is clear that both void and non-void galaxies show a main sequence of star formation. Normalized 1D probability distributions for $M_*$ and sSSF are also shown Figure~\ref{fig:ssfr} (top panels and side panel, respectively). Red crosses at $(8.38, -9.45)$ in Figure~\ref{fig:ssfr}b) and $(8.25, -9.40)$ in Figure~\ref{fig:ssfr}d) indicate the peak densities, which were determined from the relative maxima in the top and side panels.

\begin{figure*}
    \centering
    \includegraphics[width=0.95\columnwidth]{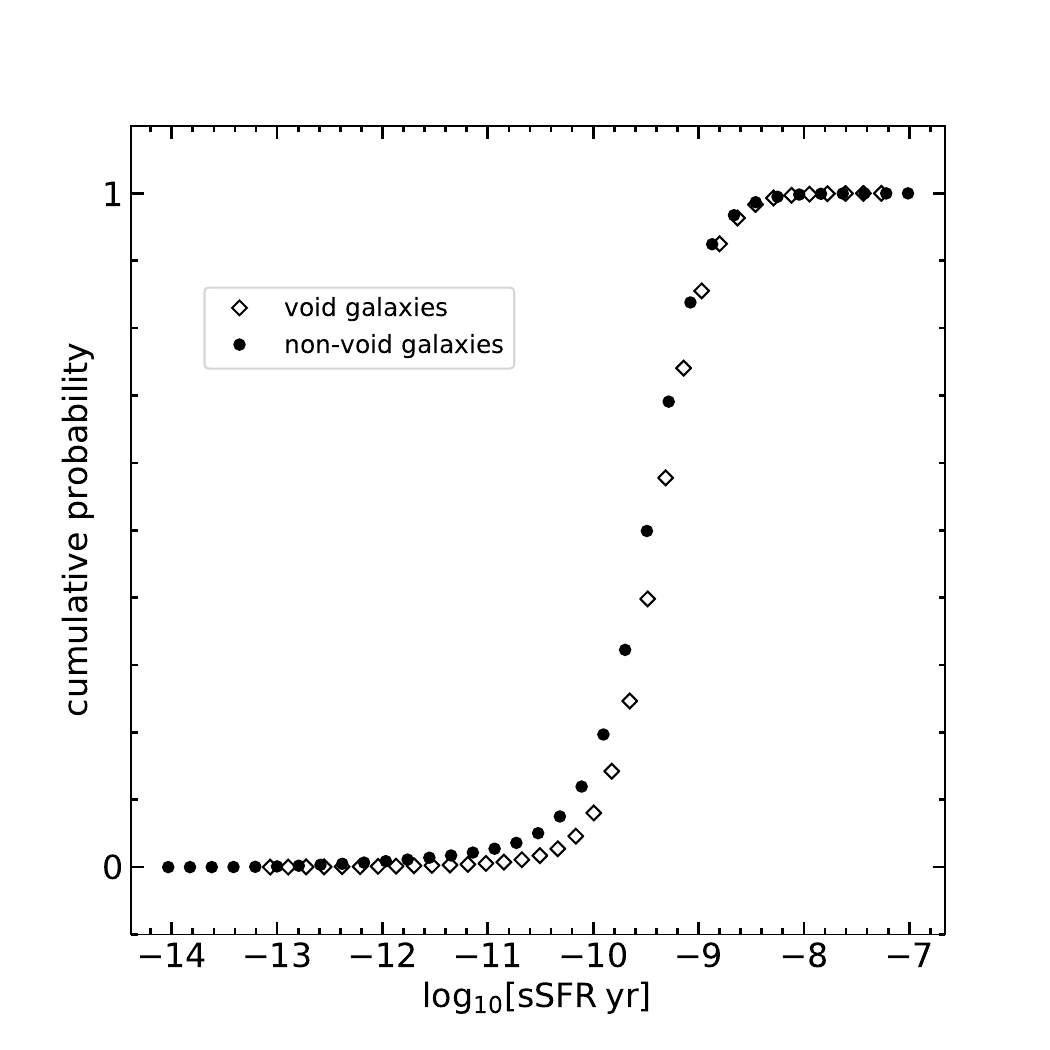}
    \caption{Normalized cumulative probability distribution functions for the specific star formation rates of void galaxies (black diamonds) and non-void galaxies (black circles).}
    \label{fig:cumulative_ssfr}
\end{figure*}

Figure \ref{fig:cumulative_ssfr} shows the normalized cumulative probability distribution functions for the sSFR of void and non-void galaxies.  A two-sample KS test shows that the two distributions are not drawn from the same underlying distribution (confidence level $> 99.9999$\%).  The median sSFR for void galaxies is $21.8 \pm 0.5$ percent higher than it is for non-void galaxies ($5.015^{+0.021}_{-0.018}\times10^{-10}\rm{yr}^{-1}$ vs.\ $4.116^{+0.008}_{-0.008}\times10^{-10}\rm{yr}^{-1}$); hence, per unit stellar mass, the void galaxies have star formation rates that are higher than those of the non-void galaxies.

\subsubsection{Supermassive Black Holes and AGN Fraction}

\begin{figure*}
    \centering
    \includegraphics[width=0.95\columnwidth]{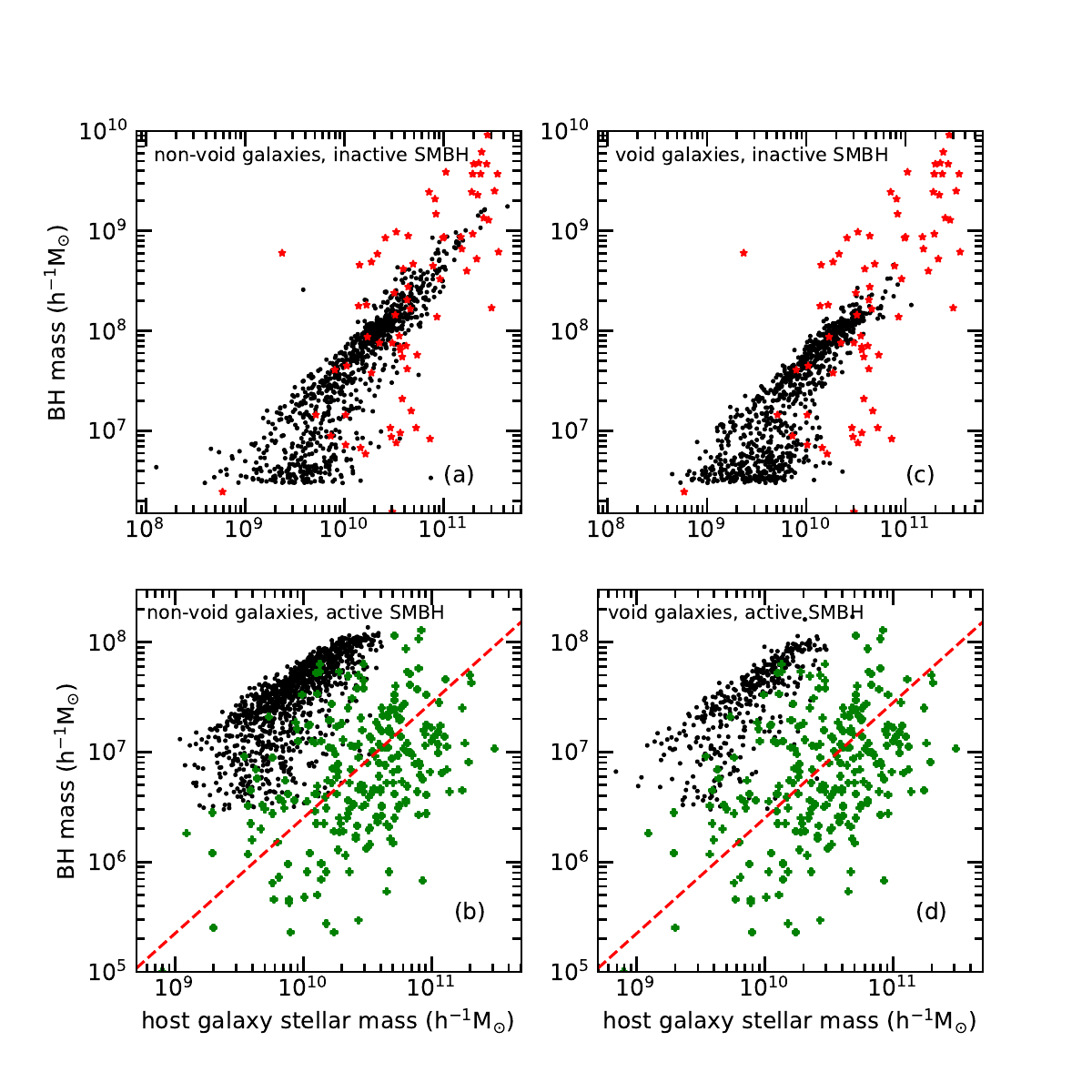}
    \caption{Black points: Supermassive black hole (SMBH) mass vs. galaxy stellar mass in the TNG300 simulation. {\it Top:} Inactive SMBH.  {\it Bottom:} Active SMBH.  {\it Left:} Non-void galaxies. {\it Right:} Void galaxies.  Red stars and green crosses: observational results obtained by \cite{reines2015}. Dashed red lines: Best-fitting relationship between SMBH mass and galaxy stellar mass for observed galaxies with AGN in \cite{reines2015}.  For clarity of the figure, randomly-selected fractions of the TNG300 data points are plotted (1\% in panel a, 10\% in panel b, 10\% in panel c, and 30\% in panel d).}
    \label{fig:smbh}
\end{figure*}

Finally, we investigate the relationships between stellar mass and supermassive black hole (SMBH) mass for TNG300 void and non-void galaxies, results of which are shown in Figure~\ref{fig:smbh} (black points). SMBHs are ``seeded'' into TNG subhalos once the subhalos pass a given mass threshold, and this seeding of SBMHs gives rise to unphysical artifacts in the stellar mass--SMBH mass relationship for SMBH with masses below $\sim 3\times 10^6 h^{-1} M_\odot$ (i.e., due to recently-seeded SMBHs having nearly identical masses, independent of the stellar mass of their host galaxy).  Therefore, we focus our analysis below on galaxies for which the masses of the SMBHs are $\gtrsim 3\times 10^6 h^{-1} M_\odot$, which results in a sample of 107,527 non-void galaxies and 10,231 void.  We further subdivide these particular galaxies according to whether or not their SMBHs are in an active state.  To do this, we use the ratio of Bondi accretion rate to Eddington accretion rate as a measure of nuclear activity.  Following \cite{weinberger2017}, we classify galaxies with ratios greater than 0.05 as ``active''.  From this classification, we find 1,330 void galaxies with SMBH masses $\gtrsim 3\times 10^6 h^{-1} M_\odot$ are in an active state (i.e., AGN fraction of $13.0 \pm 0.4$\%) and 11,269 non-void galaxies with SMBH masses $\gtrsim 3\times 10^6 h^{-1} M_\odot$ are an active state (i.e., AGN fraction of $10.5\pm 0.1$\%).  That is, for TNG300 galaxies with SMBHs in the mass range that we consider, void galaxies have a somewhat higher ($24 \pm 4$\%) AGN fraction than do non-void galaxies. We also note that it is primarily intermediate-mass SMBHs that are in an active state, since there are few TNG300 AGNs with SMBH masses $ \gtrsim 10^{8}h^{-1}$ $ M_\odot$ or $ \lesssim 10^{6.3}h^{-1}$ $ M_\odot$. 

The top panels in Figure~\ref{fig:smbh} show the relationships for galaxies with inactive SMBH, while the bottom panels show the relationships for galaxies with active SMBH.  Results for non-void galaxies are shown in the left panels of Figure~\ref{fig:smbh} and results for void galaxies are shown in the right panels.  The observed relationships between stellar mass and SMBH mass from \cite{reines2015} are also shown in Figure~\ref{fig:smbh} for comparison (red stars: inactive SMBH; green crosses: active SMBH).  Observational results for galaxies with inactive SMBH come from a sample of galaxies with dynamical BH masses (Table~3 of \citealt{reines2015}), and observational results for galaxies with active SMBH come from a sample of broadline AGN (Table~1 of \citealt{reines2015}).

From Figure~\ref{fig:smbh}, the relationship between SMBH mass and stellar mass for TNG galaxies with inactive BHs is in rough agreement with the observational results from \cite{reines2015}.  For galaxies with inactive BHs that have masses $\gtrsim 10^7 h^{-1} M_\odot$, there is a tighter relationship between SMBH mass and galaxy stellar mass in the TNG300 galaxies than there is for observed galaxies.  In contrast to observed galaxies, there are no TNG300 non-void galaxies with inactive black holes with masses $> 2\times 10^9 h^{-1} M_\odot$ and no TNG300 void galaxies with inactive black holes with masses $> 7\times 10^8 h^{-1} M_\odot$.  In the case of non-void galaxies, inactive SMBHs with masses $> 2\times 10^9 h^{-1} M_\odot$ do exist in the TNG300, but all of these objects are located within galaxies that reside in large clusters of galaxies (and all of which were omitted from our sample since the focus of our investigation is the field galaxy population).
%Also, in contrast to observed void galaxies, there are no TNG300 void galaxies with stellar masses $> 2\times 10^{11} h^{-1} M_\odot$ or SMBH masses $> 7\times 10^8 h^{-1} M_\odot$. 
%\textcolor{blue}{The maximum SMBH mass for a void galaxy is $7\times10^{8}h^{-1}M_{\odot}$. The maximum stellar mass for a void galaxy is $1.2\times10^{11}h^{-1}M_{\odot}$. So the whole statement is true, but the stellar mass statement should be changed to "with stellar masses $> 2\times10^{11}h^{-1}M_{\odot}$ if you would like both parts of the and statment to be true.}

Similar to the galaxies with inactive SMBH, TNG300 galaxies with active BHs show a much tighter relationship between stellar mass and SMBH mass than do the observed galaxies from \cite{reines2015}.  However, while the slope of the relationship is similar to that of observed galaxies with AGN, the amplitude of the relationship for TNG300 galaxies with AGN is significantly higher than it is for observed galaxies.  As a result, at fixed stellar mass, the SMBHs in TNG300 active galaxies are a factor of $\sim 10$ more massive than would be expected based on the best-fitting relationship from \cite{reines2015} (i.e., red dashed lines in Figure~\ref{fig:smbh}).

%Finally, we investigate the fraction of active galactic nuclei (AGN) within non-void and void galaxies. We classify a SMBH as an AGN if the SMBH has a Bondi accretion rate to Eddington accretion rate ratio that is greater than 0.05 (see, \citealt{weinberger2017}). Of the 198,459 non-void galaxies with SMBH masses $ >10^{6}h^{-1}$ $ M_\odot$ we find that 11,572 were in their active state which gives an AGN fraction of 5.83\%. Of the 23,630 void galaxies with SMBH masses $ >10^{6}h^{-1}$ $ M_\odot$ we find that 1,364 were in their active state which gives an AGN fraction of 5.77\%. Panels (b) and (d) show the host galaxy stellar mass vs. SMBH mass for AGN in our non-void and void galaxy populations, respectively. We note that it is mostly intermediate mass SMBH in TNG300 that are in their active state, as there are few AGN with masses $ \gtrsim 10^{8}h^{-1}$ $ M_\odot$ or $ \lesssim 10^{6.3}h^{-1}$ $ M_\odot$. \textcolor{blue}{Green plus signs in these panels show the observed relationship between SMBH mass and host galaxy stellar mass for AGN from \cite{reines2015}, and the dashed red line is their line of best fit for these data.}

\section{Summary \& Discussion}\label{sec:discussion}

Here we have investigated the properties of voids and void galaxies in the $z=0$ snapshot of the cosmological MHD simulation TNG300. The large volume and high spatial resolution of the TNG300 makes it possible to study a substantial number of voids and void galaxies.

Voids were identified using a spherical void finding algorithm that was applied to the TNG300 galaxy catalog (i.e., in analogy to observational studies of voids and void galaxies, here the voids were identified via underdensities in the distribution of luminous galaxies, not the distribution of dark matter mass or dark matter halos).  From this, a total 5,078 voids with radii that range from $2.5 h^{-1}$~Mpc to $24.7 h^{-1}$~Mpc were identified.  The median radius of the TNG300 voids is $4.4h^{-1}$~Mpc, in good agreement with the typical sizes of voids that have been found in {TNG300 \citep{davila2023} and in} previous simulations that have computational volumes similar to that of the TNG300 (see, e.g., \citealt{paillas2017}; \citealt{habouzit}).

As expected (see, e.g., \citealt{sheth}), the radial underdensity profiles of the TNG300 voids follow a reverse spherical top-hat profile.  This is the case whether luminous tracers of the underdensity (i.e., luminous galaxies) or dark matter particles are used to compute the profiles. 

%\textcolor{orange}{OC Notes: I am adding a paragraph break here. I will talk about the schuster Magneticum and DES void profile results here. The next paragraph is focused on the bias of luminous material in voids. New text is in the following blue block.} 

Recently, \cite{schuster2023} performed an in-depth study of void profiles in the \texttt{Magneticum}\footnote{\url{http://www.magneticum.org/}} suite of cosmological MHD simulations (see, e.g., \citealt{dolag2016} and \citealt{hirschmann2014} for other studies that have used \texttt{Magneticum}.) The authors find the density profiles of isolated voids to be similar for various void sizes, spatial resolutions, and mass scales. Their stacked, isolated, DM void profiles are similar to ours, although our DM profile (see Figure \ref{fig:avg_rad_profiles}) has a flatter interior out to $r=r_v$. However, since we do not investigate void profiles as functions of void shape or size, it is difficult to make direct comparisons. Regardless, the similarities to our galaxy number density profile corroborate their conclusion that the physical properties of voids are universal characteristics that are independent of tracer type and resolution. 

{\cite{davila2023} obtained integrated galaxy number density profiles for ``voids-in-clouds'' and ``voids-in-voids'' (``S-type'' and ``R-type'' voids in their notation) in TNG300. To obtain their void catalog, \cite{davila2023} used the void finding algorithm of \cite{ruiz2015}, a modified version of the void finding algorithm that we employed, but with a few differences. \cite{davila2023} considered candidate void centers as Voronoi cells constructed in the subhalo field that had density contrasts $<-0.8$. \cite{davila2023} also adopted a stricter integrated underdensity contrast threshold of $\Delta_{\rm{void}} \leq -0.9$ for their candidate voids. In addition, \cite{davila2023} did not allow for any overlap between voids; instead, they rejected all spheres that overlapped already established, larger voids. Because of this, \cite{davila2023} identify far fewer voids in TNG300 than we do here (82 voids vs.\ 5,078 voids), and the radii of their voids are restricted to a much narrow range than the radii of the voids in this work ($7 - 11 h^{-1}$~Mpc vs.\ $2.5 - 24.7 h^{-1}$~Mpc). Because of the differences in void finding algorithms and size ranges of the resulting voids, we would expect some differences between our results and those of \cite{davila2023} and, indeed, this is the case when we compare the two catalogs of ``voids-in-clouds''. In particular, the ridges of profiles for ``voids-in-clouds'' in \cite{davila2023} only reach maxima up to $\sim0.3$, which is considerably lower than the maximum for our median profile (i.e., $0.6$). However, the profiles of ``voids-in-voids'' in \cite{davila2023} are similar to the luminous profiles that we find for our ``voids-in-voids''.} 

In the observed {Universe}, \cite{sanchez2017} provide 2D {photometric} galaxy number density profiles for voids in the Dark Energy Survey (DES; \citealt{flaugher2005}; \citealt{DES}). These voids were found with a circular void finder, and utilized projected density maps of various thicknesses. Much like our voids, the interiors of DES voids are nearly empty of galaxies. However, the DES voids do not have flat interiors and gradually increase in density out to the ridges of voids. It is unclear whether this is due to projection effects caused by using 2D spectroscopic slices. %\textcolor{red}{Similarly, \cite{mao2017} provide void catalogs of voids in the SDSS Baryon Oscillation Spectroscopic Survey (BOSS; \citealt{BOSS}) Data Release 12. Their void profiles are similar to those of \cite{sanchez2017}, in that they have central densities that are $30\%$ the mean of the sample with profiles that rise gradually starting at $\sim0.2R_{\rm eff}$ to a maximum density that is $20\%$ greater than the mean of the sample.}

{Furthermore, several authors have provided void-galaxy cross-correlation functions for large voids in various SDSS, SDSS Baryon Oscillation Spectroscopic Survey (BOSS; \citealt{BOSS}), and SDSS extended BOSS (eBOSS; \citealt{eBOSS}) Data Releases (see, e.g., \citealt{nadathur2019}; \citealt{hamaus2020}; \citealt{woodfinden2022}). For instance, \cite{nadathur2019} and \cite{woodfinden2022} report voids with very empty and flat interiors ($\xi_0\sim-1.0$) out to around $20h^{-1}$Mpc that reach maxima of $\xi_0\sim0.05$ at $\sim60R_v$. In addition, \cite{hamaus2020} report values of $\xi_0$ that rise from $\sim-0.90$ in the innermost regions to $-0.55$ by $0.5R_v$ and $0.2$ at $1R_v$. Thus, the innermost regions of our TNG300 voids appear to have a higher galaxy density than those in the observed Universe, but their profiles rise more gradually out to the effective radii of voids. Despite this, the ridges of TNG300 voids have a higher overdensity of galaxies not seen in the observed Universe.}

{In terms of the detectability of our voids, the small size of most of our voids means their average galaxy number density is orders of magnitudes larger than what is available in most current observational surveys. For instance, we report an average galaxy number density of TNG300 voids of $9.4\times10^{-2}h^3\rm \: Mpc^{-3}$, whereas \cite{mao2017} found an average galaxy number density of $3.6\times10^{-4}h^3\rm \: Mpc^{-3}$ in their SDSS BOSS DR12 void member catalogs.
%and \cite{sanchez2017} have an average 2D number density of $\sim9\times10^{-4}h^2\rm \: Mpc^{-2}$ for DES voids. 
Because of this, it is unlikely that the smaller voids we report would be detected in current-generation surveys.}

We find that the TNG300 voids are more devoid of galaxies than they are of dark matter, and this result is independent of the local background density within which the voids are embedded (i.e., voids-in-voids vs.\ voids-in-clouds). That is, within the voids, mass does not trace light.  
%\textcolor{red}{has anybody else looked at void profiles in simulations and compared the profiles based on galaxies to profiles based on the dark matter? if so, we need to compare and contrast here} \textcolor{orange}{OC Notes: I have found two good articles that have directly compared luminous tracers to the underlying dark matter field: \cite{Ricciardelli2014} and \cite{pollina2017}. The former uses SDSS galaxies and the Masclet simulation. The latter used data from the Magneticum Pathfinder simulation and specifically tested the linearity of bias of luminous tracers compared to the dark matter field. Figure 3 in this paper shows that inside voids, there is a linear relationship between the density contrast of tracers and the density contrast of dark matter when galaxies, clusters, or AGN are used as tracers. I'm not calculating the bias factor myself (although I could if you think it would be beneficial), but there does seem to be a linear shift in the interior portions of our profiles in Figure 2. I have summarized this in a few sentences in the following blue text.} 
This result agrees with previous studies that have compared luminous tracers within voids to the underlying dark matter field (see, e.g., \citealt{Ricciardelli2014}; \citealt{pollina2017}). In particular, \cite{pollina2017} used the \texttt{Magneticum} suite to test the linearity of the bias of luminous tracers compared to the underlying dark matter field. There, the authors found a linear relationship between the density contrast of tracers and the density contrast of dark matter within voids when galaxies, clusters, or AGN are used as tracers. 

In addition, we note that previous work on the locations of satellite galaxies in the Illustris-1 \citep{illustris1, illustris2, illustris3, illustris4} and TNG100 
%\textcolor{orange}{(OC notes: I've added citations for Illustris-1)} 
simulations have demonstrated that mass does not trace light in host-satellite systems (see, e.g., \citealt{brainerd2018}; \citealt{mcdonough2022}). Hence, the distribution of luminous galaxies in a $\Lambda$CDM universe is an unreliable tracer of the dark matter distribution in overdense regions of space. {However, the linear bias between the galaxy and dark matter distributions in underdense regions still makes galaxies valuable probes of the density field within the linear regime of voids.} %Hence, the distribution of luminous galaxies in a $\Lambda$CDM universe is an unreliable tracer of the dark matter distribution in both overdense and underdense regions of space.

In TNG300 we identified a total of 75,220 void galaxies and 527,454 non-void field galaxies, and systematic differences between the void and non-void galaxies are clear.  Compared to the non-void galaxies, the void galaxies are, on average, younger, bluer in color, less metal enriched, have lower stellar masses, and have smaller physical extents.  The luminosity functions of both void and non-void galaxies exhibit similar faint-end slopes, but the luminosity of an $L_\ast$ non-void galaxy is $\sim 70$\% greater than the luminosity of an $L_\ast$ void galaxy, consistent with void galaxies being smaller and less massive than non-void galaxies on average.  In addition, void galaxies have a somewhat higher specific star formation rate than non-void galaxies and, in the case of galaxies with central SMBHs with masses $\gtrsim 3\times 10^6 h^{-1} M_\odot$, void galaxies have a somewhat higher AGN fraction than non-void galaxies.
These results are in agreement with previous studies that find void galaxies to be bluer, lower in stellar mass, and less metal enriched than non-void galaxies (see, e.g., \citealt{rojas2004}; \citealt{florez2021}; \citealt{rosasguevara2022}) and that concluded AGN fraction is not strongly dependent upon the local matter density of a galaxy (see, e.g., \citealt{carter2001}; \citealt{karhunen2014}; \citealt{habouzit}).
%\textcolor{red}{note: I'm not convinced that the ANG fraction is related to lack of merger activity; it could be that the AGNs in voids have had more recent mergers than AGNs outside of voids -- you haven't directly investigated this, but you certainly could, and without direct investigation you're only speculating.} \textcolor{orange}{OC notes: I agree that I cannot make a statement regarding lack of mergers in this paper.}

The relationship between central SMBH mass and host galaxy stellar mass was also investigated.  In the case of TNG300 galaxies with inactive SMBHs, the SMBH mass--stellar mass relationships of void and non-void galaxies is in rough agreement with results obtained by \cite{reines2015} for observed galaxies.  However, for galaxies with SMBH masses $\gtrsim 10^7 h^{-1} M_\odot$, the SMBH mass--stellar mass relationship is considerably tighter for TNG300 galaxies than it is for observed galaxies. A relationship between SMBH mass and stellar mass that is tighter for TNG300 galaxies than observed galaxies is also shown by void and non-void TNG300 galaxies with active central black holes.  In both cases, the tightness of the relationship for simulated galaxies may simply be reflective of the relative ease with which both parameters can be obtained in simulation space.   Lastly, while the slopes of the SMBH mass--stellar mass relationships for active TNG300 void and non-void galaxies agree well with that of observed galaxies, at fixed stellar mass the SMBHs in active TNG300 galaxies are $\sim 10$~times more massive than would be expected based on the best-fitting relationship from \cite{reines2015}. 

{The lower host galaxy stellar masses and higher SMBH masses of TNG300 AGN compared to those of \cite{reines2015} could be caused by several factors. For example, in TNG300, a SMBH of mass $6.2\times10^6 M_\odot$ is seeded into any friends-of-friends halo whose mass exceeds $7.3\times 10^{10} M_\odot$, which is based on the host-halo relationship of \cite{dimatteo2008} and \cite{sijack2009}. This initial mass could be too large, resulting in more massive SMBHs in TNG300 by $z=0.0$. This would, however, also affect the relationships in the top panels of Figure~\ref{fig:smbh}. Another explanation could involve the AGN feedback models in TNG300, which describe how energy is injected into the galaxies and circumgalactic mediums (CGM). \cite{zinger2020} find the AGN feedback channel of TNG300 AGN to be both highly ``ejective'' and ``preventative''. This means the AGN state is highly efficient at expelling star-forming gas from the galaxy and increasing the entropy of the CGM, which strongly quenches future star formation within the galaxy. Therefore, if the feedback injection is over-tuned, or if TNG300 galaxies spend too long in their ``active'' state, this could cause in the host galaxy stellar masses of TNG300 AGN to be lower than those of \cite{reines2015}.}

Our results for the systematic differences between void and non-void galaxies in the TNG300 simulation are consistent with the expectation that the two populations of galaxies underwent somewhat different evolutionary paths, with non-void galaxies forming in regions of space that had both higher gas densities and higher galaxy densities than void galaxies.  In a universe in which structure forms hierarchically, this naturally leads to non-void galaxies forming earlier than void galaxies (i.e., due to biased galaxy formation in which the highest peaks collapse first) and becoming larger on average than void galaxies (i.e., due to both a larger local reservoir of gas and a higher frequency of galaxy collisions outside the voids).  Here we have investigated void and non-void galaxy properties in a single simulation snapshot (corresponding to the present epoch) and, therefore, we cannot make definitive statements about the degree to which the evolutionary paths of the TNG300 void and non-void galaxies differed, and the ways in which those differences affected their physical properties at $z = 0$.  Further work, concentrated on the details of the evolution of void and non-void galaxies over cosmic time, will be necessary to establish and quantify differences in the evolutionary paths and the resulting effects on the natures of the two populations of galaxies.

\section*{Acknowledgements}
{We are grateful to the anonymous reviewer for helpful comments and suggestions that improved the manuscript.} This work was partially supported by National Science Foundation grant AST-2009397. The IllustrisTNG simulations were undertaken with compute time awarded by the Gauss Centre for Supercomputing (GCS) under GCS Large-Scale Projects GCS-ILLU and GCS-DWAR on the GCS share of the supercomputer Hazel Hen at the High Performance Computing Center Stuttgart (HLRS), as well as on the machines of the Max Planck Computing and Data Facility (MPCDF) in Garching, Germany. In addition, we are pleased to acknowledge that the computational work reported on in this paper was performed on the Shared Computing Cluster which is administered by Boston University’s Research Computing Services.

%%%%%%%%%%%%%%%%%%%%%%%%%%%%%%%%%%%%%%%%%%%%%%%%%%

%\bibliographystyle{aastex631}
\bibliography{bibliography}

\label{lastpage}
\end{document}